\documentclass[prc,floatfix, twocolumn,showpacs, showkeys, preprintnumbers, nofootinbib, superscriptaddress,10pt]{revtex4-1}
\usepackage[utf8]{inputenc}
\usepackage[sort&compress]{natbib}
\usepackage{ulem}
\usepackage{bm,bbm}
\usepackage{times}
\usepackage{amssymb,amsbsy,amsmath,amsfonts}
\usepackage{graphicx}
\usepackage{float}
\usepackage[dvipsnames,svgnames,x11names]{xcolor}
\usepackage{morefloats}
\usepackage{srcltx}
\usepackage{slashed}
\usepackage{subfig}
\usepackage{multirow}
\usepackage{verbatim}
\usepackage{hyperref}
\usepackage{tabularx}
\usepackage{braket}
\usepackage{lipsum}

\def\dag{\dagger}
\def\l{\left}
\def\r{\right}
\def\T{\mathrm{T}} % transpose 
\DeclareUnicodeCharacter{3000}{HEREHEREHERE}

\begin{document}

\title{Reexamination of antinucleon-nucleon interactions in covariant chiral effective field theory  }

\author{Yang Xiao}
\affiliation{School of Space and Environment, Beihang University, Beijing 102206, China}
\affiliation{School of Physics, Beihang University, Beijing 102206, China}

\author{Jun-Xu Lu}
\affiliation{School of Physics, Beihang University, Beijing 102206, China}

\author{Li-Sheng Geng}
\email[Corresponding author: ]{lisheng.geng@buaa.edu.cn}
\affiliation{School of
Physics,  Beihang University, Beijing 102206, China}
\affiliation{Peng Huanwu Collaborative Center for Research and Education, Beihang University, Beijing 100191, China}
\affiliation{Beijing Key Laboratory of Advanced Nuclear Materials and Physics, Beihang University, Beijing 102206, China }
\affiliation{Southern Center for Nuclear-Science Theory (SCNT), Institute of Modern Physics, Chinese Academy of Sciences, Huizhou 516000, China}

\begin{abstract}
Motivated by the recent progress in developing high-precision relativistic chiral nucleon-nucleon interactions, we study the antinucleon-nucleon interaction  in a hybrid approach where the real part of the potential is constructed in the leading-order covariant chiral effective field theory, and the imaginary part is described following the procedure adopted in the heavy baryon chiral effective field theory.  The phase shifts and inelasticities with $J\leq 1$ are obtained and compared to those calculated in the next-to-leading order heavy baryon chiral effective field theory. For most partial waves, the descriptions of phase shifts and inelasticities in the hybrid approach are comparable to those in the next-to-leading order  heavy baryon chiral effective field theory, confirming the relatively faster convergence of the relativistic approach observed in the nucleon-nucleon sector. In addition, we search for bound states/resonances near the $\bar{N}N$ threshold and find several structures that can be associated with those states recently observed by the BESIII Collaboration.

\end{abstract}

\maketitle

\section{Introduction}
There has been ongoing interest in antinucleon-nucleon ($\bar{N} N$) interactions over the last decade. One primary motivation is the observations of near-threshold $\bar{N}N$ enhancements in charmonium~ decays~\cite{BES:2003aic,BES:2005ega,BESIII:2011aa,BESIII:2016fbr,BESIII:2021xoh,BESIII:2023vvr}, $B$ meson decays~\cite{Belle:2002bro,Belle:2002fay}, and $e^+e^- \rightarrow \bar{p}p$ reactions~\cite{BaBar:2005sdl,BaBar:2005pon} . Those observations provided an opportunity to elucidate the existence of speculated $\bar{N}N$ molecules and stimulated studies of the $\bar{N}N$ interactions at low energies. Other motivations include the novel proposal of a super $J/\psi$ factory~\cite{Yuan:2021yks} and the construction of next-generation facilities, such as the Facility for Antiproton and Ion Research (FAIR) in Darmstadt~\cite{Sturm:2010yit} and the Super Tau-Charm Facility (STCF) in Huizhou~\cite{Peng:2020orp}.

The experimental advances have revived theoretical studies. Early studies on the $\bar{N}N$ interactions are mainly by phenomenological models~\cite{Dover:1980pd,Dover:1981pp,Cote:1982gr,Timmers:1984xv,Hippchen:1991rr,Mull:1991rs,Mull:1994gz,Entem:2006dt,El-Bennich:2008ytt}.  Inspired by the pioneering work of Weinberg~\cite{Weinberg:1990rz,Weinberg:1991um,Weinberg:1992yk}, state-of-the-art microscopic $\bar{N}N$ interactions have been constructed based on the chiral effective field theory (ChEFT). ChEFT is an effective field theory of QCD, which satisfies all relevant symmetries of QCD for momenta below
$\Lambda_{\chi} \sim 1 $ GeV, especially the chiral symmetry and its breaking patterns, accompanied by low-energy constants (LECs) that parameterize high-energy physics. By utilizing the so-called power counting rule, the relative importance of various terms contained in the most general Lagrangians can be organized self-consistently, endowing some distinct characteristics compared to the phenomenological models, such as self-consistent incorporation of many-body interactions, systematic improvement in accuracy, and reliable estimation of theoretical uncertainties. 

Historically, Weinberg's idea was first realized in the $NN$ sector~\cite{Ordonez:1993tn,vanKolck:1994yi}. Nowadays, the chiral nuclear force has been constructed up to the fifth order~\cite{Epelbaum:2014sza,Reinert:2017usi,Entem:2017gor}, becoming the cornerstone of  \textit{ab initio} nuclear studies~\cite{Machleidt:2023jws}. The $\bar{N}N$ interaction, although remaining poorly understood compared to the $NN$ interaction because of limited experiment data and sophisticated annihilation processes, is closely connected to the $NN$ interaction in ChEFT in the sense that the intermediate/long-range part of the potential can be obtained by performing $G$-parity transformations to the pion exchange potentials. In contrast, the short-range/annihilation part is described by introducing real/complex contact terms in analogy to the $NN$ interaction with LECs adjusted to data. There are several varieties of chiral $\bar{N}N$ interactions~\cite{Chen:2011yu, Kang:2013uia, Dai:2017ont}. The most accurate chiral $\bar{N}N$ interaction to date was constructed by the J{\"u}lich group~\cite{Kang:2013uia, Dai:2017ont}. The J{\"u}lich potential has some successful applications in the studies of nucleon electromagnetic form factors~\cite{Haidenbauer:2014kja,Yang:2022qoy}, semileptonic baryonic decays~\cite{Cheng:2017qpv}, near $\bar{p}p$ threshold structures~\cite{Dai:2018tlc,Yang:2022kpm}, and neutron-antineutron oscillations~\cite{Haidenbauer:2019fyd}. However, there is a long-standing renormalization-group (RG) invariance issue rooted in the Weinberg power counting, suggesting a modification on the basic assumption of this approach, namely naive dimensional analysis (NDA)~\cite{Epelbaum:2018zli,vanKolck:2020llt,Zhou:2022loi}.  

One possible solution to the NDA is its covariant counterpart. It has long been noticed that Lorentz covariance sheds light on a variety of long-standing puzzles in the baryonic sector, such as baryon magnetic
moments~\cite{Geng:2008mf}, Compton scattering off protons~\cite{Lensky:2009uv}, pion nucleon scattering~\cite{Alarcon:2011zs},  baryon masses~\cite{MartinCamalich:2010fp,Ren:2012aj}, and the two-pole structures~\cite{Lu:2022hwm}. Motivated by these successful applications and the need for relativistic studies of nuclear structure and reactions, a relativistic chiral nuclear force based on the covariant NDA was proposed in 2018~\cite{Ren:2016jna,Xiao:2018jot} and reached the level of high precision very recently~\cite{Lu:2021gsb}. Apart from an accurate description of the $NN$ data and better convergence, the covariant framework exhibits unique advantages in improving the renormalization group invariance of the $^1S_0$~\cite{Ren:2017yvw} and $^3P_0$~\cite{Wang:2020myr} partial waves, accelerating the two-pion exchange convergence~\cite{Xiao:2020ozd,Wang:2021kos}, providing better extrapolation of the lattice QCD simulations to the unphysical regime~\cite{Bai:2020yml,Bai:2021uim},  solving the $A_y$ puzzle~\cite{Girlanda:2018xrw}, and naturally explaining the saturation of nuclear matter~\cite{Zou:2023quo}, in comparison with its non-relativistic counterparts. Encouraged by these successful applications, studying the $\bar{N}N$ interaction in the covariant ChEFT is intriguing to explore whether the aforementioned distinct features hold in the $\bar{N}N$ system. 

In this work, we construct the antinucleon-nucleon interaction in a hybrid approach in which the elastic process is determined in the leading order (LO) covariant ChEFT, and the annihilation process is described in the next-to-leading order (NLO) heavy baryon (HB) ChEFT.  A relativistic three-dimensional reduction of the Bethe-Salpeter equation is used to obtain the scattering amplitude from the hybrid potential. All 26 LECs parameterizing the short-range and annihilation potentials are fixed by fitting to the energy-dependent Nijmegen partial wave analysis (PWA)  of the $\bar{p}p$ data~\cite{Zhou:2012ui}. A satisfactory description of the phase
shifts and inelasticities of low angular momenta is achieved in analogy to the pertinent relativistic $NN$ interaction. 

The paper is organized as follows. In Sect.~\ref{sec:V_LO}, we explain how to derive the hybrid potentials. The scattering equation and the procedure to obtain the phase shifts are shown in Sect.~\ref{sec:Scat}. In Sect.~\ref{sec:result}, the $\bar{N}N$ phase shifts for $J \leq 1$ partial waves are calculated, and possible near-threshold bound/resonant states are searched for. Finally, we provide a summary in Sect.~\ref{sec:summary}. 

\section{The hybrid antinucleon-nucleon potentials }
\label{sec:V_LO}

\begin{figure*}[htbp]
\centering
\includegraphics{ 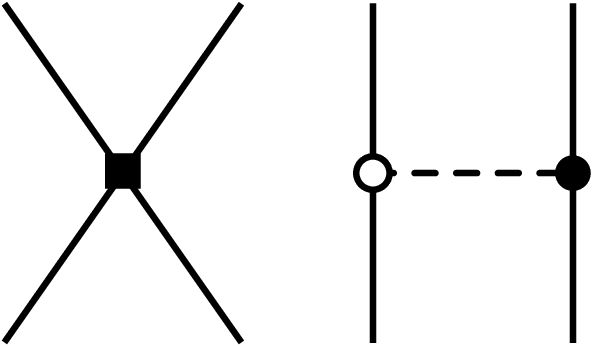}
\caption{Feynman diagrams contributing to the elastic part of the
$\bar{N}N$ interaction at leading order in the covariant power counting. The solid lines denote nucleons/antinucleons, and the dashed line represents the pion. The box denotes the vertex from $\mathcal{L}_{\bar{N}N}^{(0)}$, while the circle/dot shows vertex from $\mathcal{L}_{\pi \bar{N}}^{(1)}$/$\mathcal{L}_{\pi N}^{(1)}$.}
\label{Fig:LO}
\end{figure*}
The $\bar{N}N$ interaction contains real and imaginary parts, which reads 
\begin{align}
    V_{\bar{N}N} = V^{\text{R}} + V^{\text{I}}.
\end{align}
For the real parts of the interaction, 
the underlying covariant power counting is the same as the $NN$ case, which is described in detail in Refs.~\cite{Ren:2016jna,Xiao:2018jot,Lu:2021gsb}, where  the antinucleon field (spinor $v$) and the nucleon field (spinor $u$) are treated on an equal footing as spin-1/2 fields. The corresponding LO Feynman diagrams are summarized in Fig.~\ref{Fig:LO}, and the relevant Lagrangians are,

\begin{align}
\label{eq:lageff}
\mathcal{L}_{\text{eff.}}=\mathcal{L}_{\pi\pi}^{(2)} + \mathcal{L}_{\pi \bar{N}}^{(1)} + \mathcal{L}_{\pi N}^{(1)} + \mathcal{L}_{\bar{N}N}^{(0)},
\end{align}
where the superscript denotes the chiral dimension. The
lowest order $\pi\pi$, $\pi \bar{N}$, $\pi N$, and $\bar{N}N$   Lagrangians read,
\begin{align}
\label{eq:Lag}
\mathcal{L}_{\pi\pi}^{(2)} =& \frac{f_\pi^2}{4} \text{Tr} \left[ \partial_\mu U \partial^\mu U^\dag + \left( U + U^\dag\right)m_\pi^2 \right],\\
\mathcal{L}_{\pi \bar{N}}^{(1)}  =& \mathcal{L}_{\pi N}^{(1)} = \bar{\Psi}\left( i \slashed{D} - M + \frac{g_A}{2} \gamma^\mu \gamma_5 u_\mu \right)\Psi,\\
\mathcal{L}_{\bar{N} N}^{(0)} =&  C_S \left( \bar{\Psi} \Psi \right)\left( \bar{\Psi} \Psi \right) + C_A \left( \bar{\Psi} \gamma_5 \Psi \right) \left( \bar{\Psi} \gamma_5 \Psi \right) \\\nonumber  
  +&  C_V\left( \bar{\Psi} \gamma_\mu \Psi \right)\left( \bar{\Psi} \gamma^\mu \Psi \right) \\\nonumber
 +&  C_{AV}\left( \bar{\Psi} \gamma_\mu \gamma_5 \Psi \right) \left( \bar{\Psi} \gamma^\mu \gamma_5 \Psi \right) \\\nonumber
 +& C_T\left( \bar{\Psi} \sigma_{\mu\nu} \Psi \right)\left( \bar{\Psi} \sigma^{\mu\nu} \Psi \right), 
\end{align}
with the pion decay constant $f_\pi = 92.4 $ MeV,  the axial  coupling constant $g_A = 1.29$~\cite{Machleidt:2011zz}, the $SU(2)$ matrix $U = u^2 = \text{exp} \left(\frac{i \Phi}{f_\pi}\right)$, where $\Phi$ and $\Psi$ are,
\begin{equation}
    \Phi = \left( 
    \begin{matrix} \pi^0 & \sqrt{2} \pi^+ \\ 
    \sqrt{2}\pi^- & -\pi^0 
    \end{matrix}
    \right), ~~~~ \Psi =\left(
    \begin{matrix}
        p \\ n
    \end{matrix}
    \right).
\end{equation}
The covariant derivative of the nucleon field $\Psi$ is defined as, 
\begin{align}
    D_\mu \Psi &= \partial_\mu \psi +\left[\Gamma_\mu, \Psi\right], \\
    \Gamma_\mu & = \frac{1}{2} \left( u^\dag \partial_\mu u + u \partial_\mu u^\dag\right), 
\end{align}
and the axial current $u_\mu$ is,
\begin{align}
    u_\mu = i \left( u^\dag \partial_\mu u - u \partial_\mu u^\dag\right).
\end{align}
The LO covariant potentials $V^\text{R}_{\text{LO}}$ can be obtained by summing the contact (CT) and one-pion-exchange (OPE) terms shown in Fig.~\ref{Fig:LO},
\begin{align}
    V_\text{LO}^{\text{R}}=V^{\text{R}}_\text{CT} + V^{\text{R}}_\text{OPE},
\end{align}
where the contact potential $V^{\text{R}}_\text{CT}$ is,
\begin{widetext}
\begin{align}
\label{eq:LO_CT}
V^{\text{R}}_\text{CT}\l( \bm{p},\bm{p}'\r)= & C_S\l[ \bar{v}\l( \bm{p},s_1\r) v\l( \bm{p}' ,s_1' \r) \r] \l[ \bar{u}\l( -\bm{p}' ,s_2' \r)  u \l( -\bm{p}, s_2 \r)\r]
+  C_A\l[ \bar{v}\l( \bm{p},s_1\r) \gamma_5 v\l( \bm{p}' ,s_1' \r) \r] \l[ \bar{u}\l( -\bm{p}' ,s_2' \r) \gamma_5 u \l( -\bm{p}, s_2 \r)\r] \\\nonumber
+ & C_V\l[ \bar{v}\l( \bm{p},s_1\r)\gamma_\mu v\l( \bm{p}' ,s_1' \r) \r] \l[ \bar{u}\l( -\bm{p}' ,s_2' \r) \gamma^\mu u \l( -\bm{p}, s_2 \r)\r] 
+  C_{AV}\l[ \bar{v}\l( \bm{p},s_1\r) \gamma_\mu\gamma_5v\l( \bm{p}' ,s_1' \r) \r] \l[ \bar{u}\l( -\bm{p}' ,s_2' \r) \gamma^5\gamma_5 u \l( -\bm{p}, s_2 \r)\r] \\\nonumber
+ & C_T\l[ \bar{v}\l( \bm{p},s_1\r) \sigma_{\mu\nu} v\l( \bm{p}' ,s_1' \r) \r] \l[ \bar{u}\l( -\bm{p}' ,s_2' \r) \sigma^{\mu\nu} u \l( -\bm{p}, s_2 \r)\r],
\end{align}
and the one-pion-exchange potential $V^{\text{R}}_{\text{OPE}}$ is
\begin{align}
\label{eq:LO_OPE}
    V^{\text{R}}_{\text{OPE}}\l( \bm{p},\bm{p}' \r) = \frac{g_A^2}{4f_\pi^2}\frac{ \l[ \bar{v}\l( \bm{p},s_1 \r) \bm{\tau}_1 \gamma_\mu \gamma_5 q^\mu v\l( \bm{p}',s_1' \r) \r] \l[\bar{u}\l( -\bm{p}',s_2' \r) \bm{\tau}_2 \gamma_\nu \gamma_5 q^\nu u\l( -\bm{p},s_2 \r)\r]}{\l(E_{p'}-E_{p}\r)^2 -\l( \bm{p}' - \bm{p}\r)^2 -m_\pi^2 },
\end{align}
\end{widetext}
where $p$/$p'$ is the incoming/outgoing three momentum, $m_\pi$ refers to the pion mass and we use the isospin-averaged value $m_\pi=138$ MeV, $q^\mu=\l( E_{p'}-E_p, \bm{p}'-\bm{p} \r)$, and $\bm{\tau}$ is the isospin Pauli matrix. The Dirac spinor $u(\bm{p},s)$ is,
\begin{equation}
    u(\bm{p},s)= N_p \l(
    \begin{matrix}
        1 \\
        \frac{ \bm{\sigma} \cdot \bm{p}}{ \epsilon_p}
    \end{matrix}
    \r) \chi_s, ~~~~N_p = \sqrt{\frac{\epsilon_p}{2m_N}},
\end{equation}
where $m_N$ refers to the nucleon mass, and we use the isospin-averaged value $m_N=939$ MeV, $\epsilon_p=E_p+m_N$, $\chi_s$ denotes the Pauli spinor matrix, and $\bm{\sigma}$ is the Pauli matrix. The Spinor $v\l(\bm{p},s\r)= \gamma_0 C u^*\l( \bm{p},s \r)$ with $C$ representing the charge transformation operator,
\begin{equation}
    C = i \gamma_0 \gamma_2 =\l(
    \begin{matrix}
        0 & i \bm{\sigma}_2\\
        i \bm{\sigma}_2 &0
    \end{matrix}
    \r).
\end{equation}

The antinucleon-nucleon contact terms include one antinucleon field $v$, one nucleon field $u$, and their adjoint fields $\bar{v}$ and $\bar{u}$. The different arrangements of these four-baryon fields are
of the following schematic form:
\begin{align}
\nonumber
    &\sum_i C_i\l[ \bar{v}\l(\bm{p},s_1\r) \Gamma_i v\l(\bm{p}',s_1'\r)\r]\l[ \bar{u}\l(- \bm{p}',s_2' \r) \Gamma^i u\l( -\bm{p},s_2 \r) \r],\\\nonumber
    &\sum_i C_i\l[ \bar{v}\l(\bm{p},s_1\r) \Gamma_i u\l(-\bm{p},s_2\r)\r]\l[ \bar{u}\l( -\bm{p}',s_2' \r) \Gamma^i v\l( \bm{p}',s_1' \r) \r],\\\nonumber
    &\sum_i C_i\l[ \bar{v}\l(\bm{p},s_1\r) \Gamma_i u\l(-\bm{p},s_2\r)\r]\l[ \bar{u}\l( \bm{p}',s_1' \r) \Gamma^i v\l( -\bm{p}',s_2' \r) \r],\\\nonumber
    &\sum_i C_i\l[ \bar{v}\l(\bm{p},s_1\r) \Gamma_i v\l(-\bm{p}',s_2'\r)\r]\l[ \bar{u}\l( \bm{p}',s_1' \r) \Gamma^i u\l( -\bm{p},s_2 \r) \r],
\end{align}
where $C_{i \in \{S,A,V,AV,T\}}$ refers to the low-energy constants and $\Gamma_i$ is the corresponding Clifford algebra. Using the generalized Fierz identities~\cite{Nieves:2003in}, a product of two bilinears can be rearranged as
\begin{align}
    \bm{e}\l(1234 \r) = \bm{K}^{\l( abcd\r)}\bm{e}\l(abcd \r),
\end{align}
where $\bm{e}\l(abcd \r)$ represents an ordering of quadrilinears and $\bm{K}^{\l(abcd \r)}$ is the transformation matrix, whose explicit forms are given in Appendix~\ref{sec:Fierz}. This allows one to express all the arrangements as a linear combination of the chosen type, in our case, Eq.~\eqref{eq:LO_CT}. Here the $C_A\l[ \bar{v}\l( \bm{p},s_1\r) \gamma_5 v\l( \bm{p}' ,s_1' \r) \r] \l[ \bar{u}\l( -\bm{p}' ,s_2' \r) \gamma_5 u \l( -\bm{p}, s_2 \r)\r]$ term, which arises from the next-to-leading order potential according to Refs.~\cite{Xiao:2018jot,Lu:2021gsb}, is ascended to leading order to ensure that one can make use of the generalized Fierz identities to get rid of redundant terms in the potential.

In computing the observables, it is convenient to transform the potentials into the $LSJ$ basis, where $L$ denotes the total orbital angular momentum, $S$ is the total spin, and $J$ is the total angular momentum. The procedure for the partial wave projection is standard~\cite{Erkelenz:1971caz,Erkelenz:1974uj}. The explicit expression for the OPE potential in the $LSJ$ basis is of the opposite sign as that in the $NN$ case given in Ref.~\cite{Ren:2016jna} after partial wave projection, while the contact potentials  are of the following form,
\begin{align}
\label{eq:LO_CT_LSJ}
  \nonumber 
  &V^{\text{R}}_{1S0} = \xi \l[ C_{1S0}\l( R_p^2 +R_{p'}^2 \r) +\hat{C}_{1S0}\l( 1+R_p^2R_{p'}^2 \r)   \r] ,\\\nonumber
  &V^{\text{R}}_{3P0}  = \xi C_{3P0} R_p R_{p'},\\\nonumber
  &V^{\text{R}}_{1P1}  = 2\xi \l( C_{3S1}-3\hat{C}_{3S1} \r)R_p R_{p'},\\\nonumber
  &V^{\text{R}}_{3P1}  = \frac{4}{3}\xi\l( C_{1S0}-\hat{C}_{1S0} \r) R_p R_{p'},\\\nonumber
  &V^{\text{R}}_{3S1}  = \xi \l[ C_{3S1}\l( R_p^2 +R_{p'}^2 \r) +\hat{C}_{3S1}\l( 9 +R_p^2 R_{p'}^2 \r)\r],\\\nonumber
  &V^{\text{R}}_{3D1}  = 8\xi \hat{C}_{3S1}R_p^2 R_{p'}^2, \\\nonumber
  &V^{\text{R}}_{3S1-3D1} = 2 \sqrt{2}\xi \l( 2C_{3S1} R_p^2 + \hat{C}_{3S1} R_p^2 R_{p'}^2 \r),\\
  &V^{\text{R}}_{3D1-3S1} = 2 \sqrt{2}\xi \l( 2C_{3S1} R_{p'}^2 + \hat{C}_{3S1} R_{p}^2 R_{p'}^2 \r),
\end{align}
where $\xi = -4\pi N_p^2 N_{p'}^2$, $R_p = |\bm{p}|/\epsilon_p$, and $R_{p'} = |\bm{p'}|/\epsilon_{p'}$. The low-energy constants are  linear combinations of $C_{S,A,V,AV,T}$ of the following form,
\begin{align}
\nonumber
    & C_{1S0} = C_A + C_{AV} - 6 C_T + 3 C_V,\\\nonumber
    & \hat{C}_{1S0}=3 C_{AV} + C_S - 6 C_T + C_V,\\\nonumber
    & C_{3P0} = -2 \l(C_A - 4 C_{AV} +  C_S - 12 C_T - 4 C_V \r),\\\nonumber
    & C_{3S1} = \frac{1}{3} \l(-C_A - C_{AV} - 2 C_T + C_V \r),\\
    & \hat{C}_{3S1} = \frac{1}{9} \l(-C_{AV} + C_S + 2 C_T + C_V\r).
\end{align}
The covariant contact potential contributes to all $J=0,1$ partial waves with 5 independent rearranged low-energy constants $C_{1S0}, \hat{C}_{1S0}, C_{3S1}, \hat{C}_{3S1}$, and $C_{3P0}$. The $^1P_1,^3P_1,^3D_1,^3S_1-^3D_1$, and $^3D_1-^3S_1$ potentials are constrained only by the $S$-wave parameters, which allow us to check the relativistic corrections to the short-range $\bar{N}N$ interaction. The Pauli principle does not hold in the $\bar{N}N$ interaction, so the number of low-energy constants is twice that of the $NN$ case. 

Compared to the $NN$ interaction, a new feature of the $\bar{N}N$ interaction is the presence of the annihilation process, which leads to an intrinsic difficulty in describing a system that has hundreds of annihilation many-body channels at rest~\cite{Carbonell:2023onq}. Here, we follow the approach of Ref.~\cite{Kang:2013uia} that manifestly fulfills unitarity and considers the contributions to the potential from the annihilation process of the following form,
\begin{align}
\label{eq:Vann}
V=\sum_{X=2\pi,3\pi,...}V_{\bar{N}N \rightarrow X}G_XV_{X \rightarrow \bar{N}N},
\end{align}
where $X$ is the sum over all open annihilation channels, and $G_X$ is the propagator of the
intermediate state $X$. Making use of the identity 
\begin{align}
    \frac{1}{x \pm i \epsilon} = \mathcal{P}\frac{1}{x} \mp i\pi \delta\l(x\r),
\end{align}
The imaginary part of Eq.~\eqref{eq:Vann} is constrained by,
\begin{align}
    \text{Im} V = -\pi \sum_X V_{\bar{N}N \rightarrow X}V_{X \rightarrow \bar{N}N}.
\end{align}

Expanding $V_{
\bar{N}N\rightarrow X}$ in powers of the nucleon three momentum up to NLO , one can obtain the imaginary parts of the $\bar{N}N$ interaction,
\begin{align}
\label{eq:ann_LO}
\nonumber
    &V_{1S0}^{\text{I}} = -i\l( C_{1S0}^{a} + \hat{C}_{1S0}^{a} \frac{p^2}{4m_N^2} \r)\l( C_{1S0}^{a} + \hat{C}_{1S0}^{a} \frac{p'^2}{4m_N^2} \r),\\\nonumber
    &V_{3P0}^{\text{I}} = -i \l(C_{3P0}^{a}\r)^2 \frac{pp'}{4m_N^2},\\\nonumber
    &V_{1P1}^{\text{I}} = -i \l(C_{1P1}^{a}\r)^2 \frac{pp'}{4m_N^2},\\\nonumber
    &V_{3P1}^{\text{I}} = -i \l(C_{3P1}^{a}\r)^2 \frac{pp'}{4m_N^2},\\\nonumber
    &V_{3S1}^{\text{I}} = -i\l( C_{3S1}^{a} + \hat{C}_{3S1}^{a} \frac{p^2}{4m_N^2} \r)\l( C_{3S1}^{a} + \hat{C}_{3S1}^{a} \frac{p'^2}{4m_N^2} \r),\\\nonumber
    &V_{3S1-3D1}^{\text{I}} = -i \l(C_{3S1}^{a} + \hat{C}_{3S1}^{a} \frac{p^2}{4m_N^2}\r)C_{\epsilon_1}^a \frac{p'^2}{4m_N^2},\\\nonumber
    &V_{3D1-3S1}^{\text{I}} = -i C_{\epsilon_1}^a \frac{p^2}{4m_N^2}\l(C_{3S1}^{a} + \hat{C}_{3S1}^{a} \frac{p'^2}{4m_N^2}\r),\\
    &V_{3D1}^{\text{I}} = -i \l(C_{\epsilon_1}^{a}\r)^2 \frac{p^2p'^2}{16m_N^4}.
\end{align}
The factors $\frac{1}{4m_N^2}$ and $\frac{1}{16m_N^4}$ are introduced to ensure that all annihilation constants are of the same dimension. 

There are several issues to address regarding the contributions to the potential from the annihilation process. 1) Eq.~\eqref{eq:ann_LO} only contains the $J=0,1$ partial waves to be consistent with the real parts of the potential given in Eq.~\eqref{eq:LO_CT_LSJ}. 2)  Eq.~\eqref{eq:ann_LO} is organized in the conventional Weinberg power counting, while a more self-consistent potential should be constructed in the covariant power counting. The main difficulty in evaluating a covariant potential relevant to the annihilation process is the complexity of the explicit expressions for all open annihilation channels' potentials $V_{\bar{N}N \rightarrow X}$. Based on the experience in the $NN$ interaction, the accuracy of the covariant potential is comparable to the HB potential at one order higher. Therefore, we expand $V_{\bar{N}N \rightarrow X}$ up to NLO in the conventional Weinberg power counting to evaluate the imaginary parts of the potential as an approximation of the exact covariant potential at LO.  3) The complete contributions to the potential from the annihilation process contain a real part from the principal value in Eq.~\eqref{eq:Vann},  whose structure is accounted
for by the real parts of the LECs in the conventional chiral potential at the corresponding order.  By contrast, the contribution of the real part of Eq.~\eqref{eq:Vann} can only be absorbed in the covariant LECs partly in our case because the number of independent LECs in the covariant power counting at LO is less than that in the conventional Weinberg  power counting at NLO. However, this real part does not break unitarity.  In addition, the contribution to the $\bar{N}N$ interaction from the additional structures is suppressed because it is of order $\mathcal{O}\l( p^2 \r)$ (here $p$ refers to the small quantity in the conventional Weinberg power counting). Therefore, we use the pure imaginary potential in Eq.~\eqref{eq:ann_LO} to account for the annihilation process in practice. It should be mentioned that the problem above can be solved by constructing a self-consistent potential responsible for the annihilation process in the covariant power counting. We will explore how to implement this idea in the future.

\section{Scattering equation and phase shifts}
\label{sec:Scat}

The partial wave projected scattering $T$-matrix is obtained by solving the Kadyshevsky equation in the $LSJ$ basis,
\begin{align}
\nonumber
    T^{SJ}_{L',L}\l(p',p \r)&=V^{SJ}_{L',L}\l(p',p \r) +\sum_{L''}\int_0^{+\infty} \frac{k^2 \text{d}k}{\l(2\pi\r)^3}V^{SJ}_{L',L}\l(p',k \r) \\\nonumber 
    & \times \frac{m_N^2}{2\l( k^2+m_N^2 \r)} \frac{1}{\sqrt{p^2+m_N^2}-\sqrt{k^2+m_N^2}+i\epsilon}\\
    & \times T^{SJ}_{L'',L} \l( k,p \r).
\end{align}
We separately solve the Kadyshevsky equation in the isospin basis for $I=0$ and $I=1$ and fit the resulting phase shifts to those in Ref.~\cite{Zhou:2012ui} to determine the corresponding LECs. To remove the ultraviolet divergences, the potential is regularized with a non-local Gaussian-type cut-off function,
\begin{align}\label{eq:cutoff}
    f^\Lambda \l( p,p' \r) = \exp\l[-\l( p^6+p'^6 \r)/\Lambda^6\r],
\end{align}
with the cut-off value varied in the range $\Lambda=450-600$ MeV. We want to emphasize that the employed regulator here is sort of ``old-fashioned" in the light of the studies on the $NN$ interactions in the conventional ChEFT because it distorts the long-range part of the potential thus slowing down the convergence~\cite{Reinert:2017usi}. In our framework, although the covariant OPE potential, in contrast to its conventional counterpart, is non-local due to the presence of Dirac spinors, the employed regulator might still introduce some cut-off artifacts because such a regularization implies including a sequence of terms proportional to $\frac{1}{q^2 - m_\pi^2} \left(-\frac{p^6+p'^6}{\Lambda^6}\right)^n$ in the conventional expansion for nucleons in the unregularized covariant OPE potential. Thus, more careful studies are necessary to understand a proper regularization method in our framework comprehensively. Note that the semi-local regulator has been proven to maintain the long-range part of the pion-exchange potential and remove cut-off artifacts simultaneously, as suggested in Ref.~\cite{Reinert:2017usi}. However, implementing such a regulator in the present framework is not straightforward because it involves recalculating the covariant pion-exchange potential and projecting it into the $LSJ$ basis. Additionally, such regularization is not expected to change our results significantly, which will be left for future work.  

The partial wave $S$ matrix is related to the on-shell 
$T$ matrix by,
\begin{align}
    S^{SJ}_{L',L}\l( p \r) = \delta_{L',L} - i \frac{ p~ m_N^2 }{8\pi^2 E_p} T^{SJ}_{L',L}\l(p \r).
\end{align}

Phase shifts and mixing angles can be obtained from the matrix $S$ using the idea of “Stapp”~\cite{Stapp:1956mz}. The annihilation process makes the phase shifts complex for the $\bar{N}N$ interaction. We follow the procedure of Ref.~\cite{refId0} to evaluate the phase shifts. For uncoupled channels, the real and imaginary parts of the phase shift $\delta_L$ can be obtained from the on-shell $S$ matrix,
\begin{align}
\nonumber
    \text{Re} \l(\delta_L\r) &= \frac{1}{2} \arctan \frac{\text{Im}\l(S_L\r)}{\text{Re}\l(S_L\r)}, \\
    \text{Im}\l( \delta_L \r)& = -\frac{1}{2}\log\lvert S_L  \rvert.
\end{align}
For coupled channels, the phase shifts $\delta_{L \pm1}$ and mixing angles $\epsilon_J$ are,
\begin{align}
\nonumber
 \text{Re}  \l(\delta_{L \pm 1}\r) &= \frac{1}{2} \arctan \frac{\text{Im} \l( \eta_{L \pm1} \r) }{\text{Re} \l( \eta_{L \pm1} \r) } ,\\  \nonumber
 \text{Im} \l( \delta_{L \pm 1} \r) &= -\frac{1}{2} \log \lvert \eta_{L \pm1} \rvert , \\\nonumber
  \epsilon_J &=   \frac{1}{2} \arctan \l( \frac{i \l( S_{L-1, L-1}+S_{L+1, L+1}\r)}{2\sqrt{S_{L-1, L-1} S_{L+1, L+1}}}\r),
\end{align}
where $\eta_L =  \frac{S_{L,L}}{\cos 2 \epsilon_J}  $.

\section{Results and discussions}
\label{sec:result}

In the fitting procedure, we perform a simultaneous fit to the $J = 0, 1$ PWA of Ref.~\cite{Zhou:2012ui} at laboratory energies below 125 MeV $\l( p_{\text{lab}} \leq 500  ~ \text{MeV}\r)$  with cutoff values varying in the range $\Lambda=450-600$ MeV, except for the $^{1}S_0$ and $^{3}P_0$ partial waves with $I=1$, where we consider extra data at $p_{\text{lab}}=600$ MeV because of the resonance-like behaviors. Table~\ref{tb:LECs} lists the numerical values of the LECs. The values for $\hat{C}_{3S1}$ are of one or two orders of magnitude smaller than $C_{3S1}$. Still, its contribution to the $^3S_1$ potential is comparable to that from $C_{3S1}$ because the contribution to the $^3S_1$ potential from $C_{3S1}$ is suppressed by $1/\l(4m_N^2\r)$ to some extent since it is multiplied by $R_{p}^2,R_{p'}^2$. A similar situation occurs in the ${^1S_0}$ partial wave and the annihilation process.  
\begin{table}[h!]
\centering
\caption{Low-energy constants for different cutoffs. The parameters related to the elastic  process are in units of $10^4$ GeV$^{-2}$, while the parameters related to the annihilation process are in units of $10^2$ GeV$^{-1}$ }
\label{tb:LECs}
\begin{tabular}{cc|c|c}
\hline
\hline
 \multicolumn{2}{c|}{LEC}    & $\Lambda=450$ MeV & $\Lambda=600$ MeV\\
\hline
 \multirow{13}{*}{$I=0$}&$C_{1S0}$& $0.213$ & $0.154$\\
 &$\hat{C}_{1S0}$& $0.031$& $0.013$ \\
 &$C_{1S0}^a$&$-1.080$& $0.668$\\
 &$\hat{C}_{1S0}^a$&$20.987$& $-9.199$\\
 &$C_{3P0}$&$-0.019$&$-0.117$\\
 &$C_{3P0}^a$&$1.472$&$0.971$\\
 &$C_{1P1}^a$&$1.281$&$1.478$\\
 &$C_{3P1}^a$&$0.737$&$0.447$\\
 &$C_{3S1}$&$-0.043$&$-0.025$\\
 &$\hat{C}_{3S1}$&$0.001$&$0.0002$\\
 &$C_{3S1}^a$&$0.207$&$-0.388$\\
 &$\hat{C}_{3S1}^a$&$4.533$&$3.326$\\
 &$C_{\epsilon 1}^a$&$-1.982$&$0.743$\\
 \hline
\multirow{13}{*}{$I=1$}&$C_{1S0}$& $-0.051$ & $0.016$ \\
 &$\hat{C}_{1S0}$& $-0.004$ &  $0.025$\\
 &$C_{1S0}^a$& $-0.398$ & $1.270$\\
 &$\hat{C}_{1S0}^a$& $4.730$ &  $-14.110$\\
 &$C_{3P0}$& $0.242$ & $0.179$ \\
 &$C_{3P0}^a$& $1.177$ & $0.570$\\
 &$C_{1P1}^a$& $1.270$& $1.066$\\
 &$C_{3P1}^a$& $1.336$ & $1.214$\\
 &$C_{3S1}$& $0.014$& $0.032$\\
 &$\hat{C}_{3S1}$& $0.001$ & $0.001$\\
 &$C_{3S1}^a$&$0.211$ & $-0.081$\\
 &$\hat{C}_{3S1}^a$& $9.208$& $-4.512$\\
 &$C_{\epsilon 1}^a$& $1.720$ & $-2.078$\\
\hline
\hline
\end{tabular}
\end{table}

\begin{figure*}[htbp]
\centering
\subfloat{
\includegraphics[width=0.4\textwidth]{ 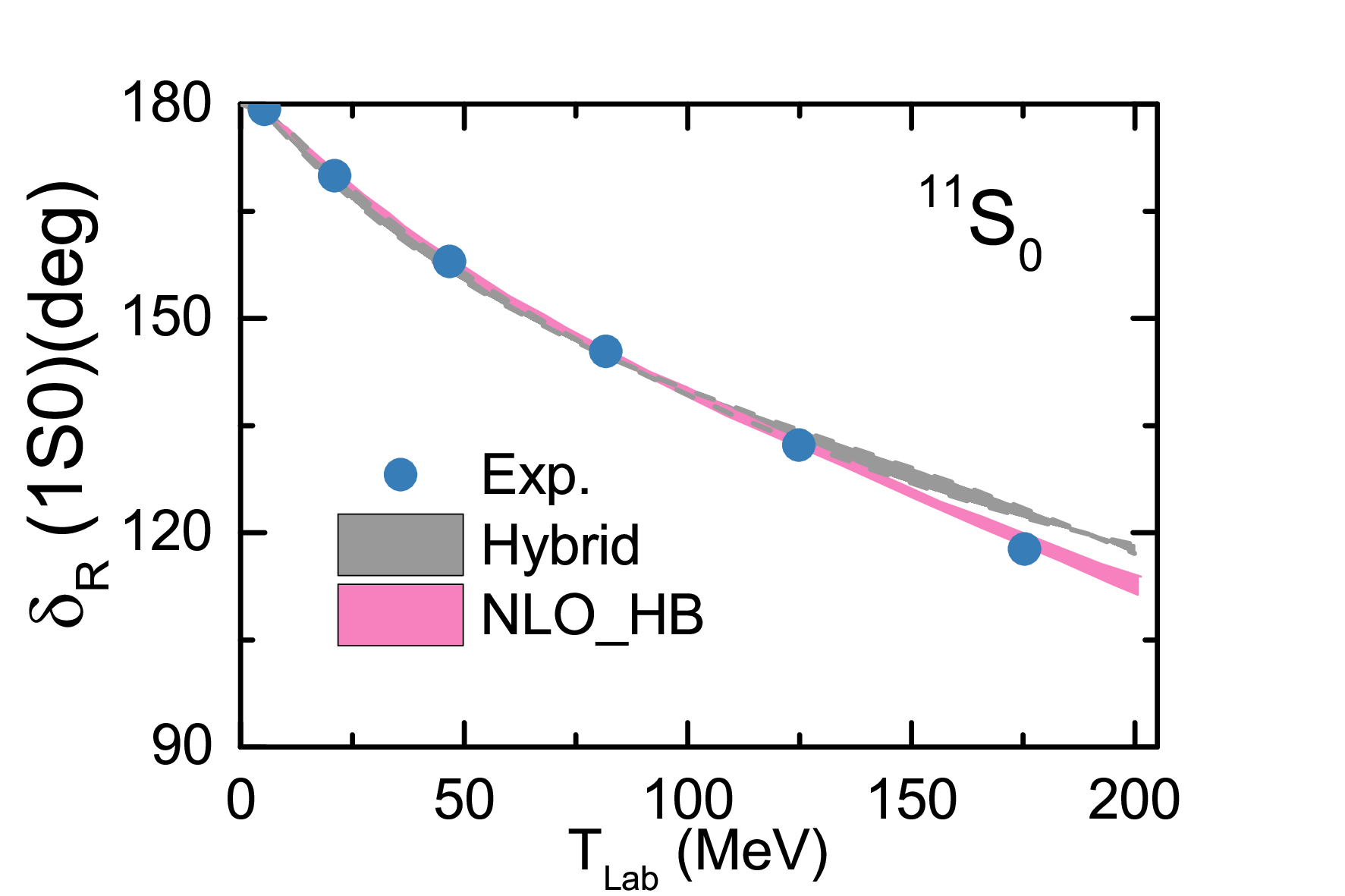}
}
\quad
\subfloat{
\includegraphics[width=0.4\textwidth]{ 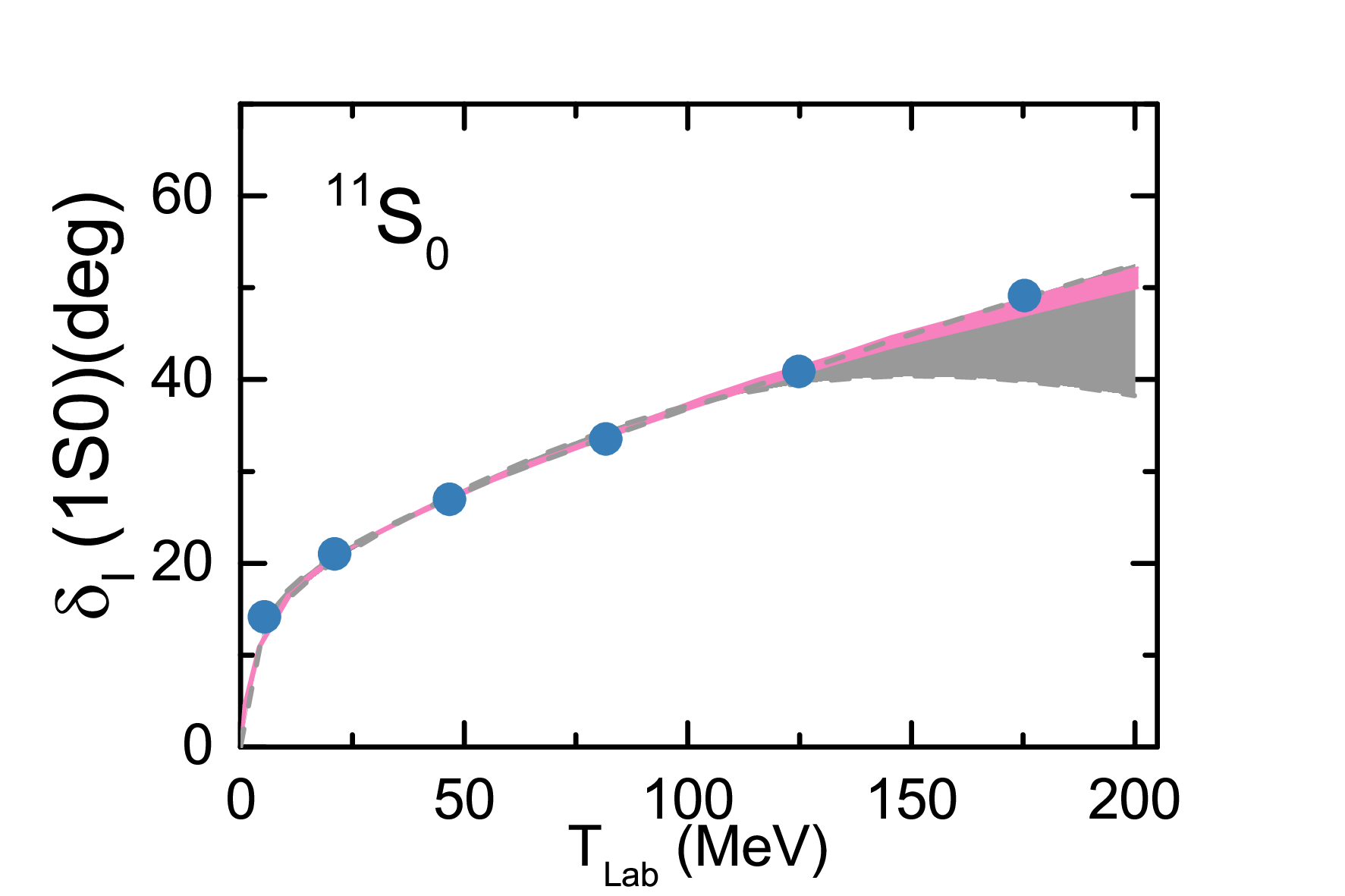}
}
\quad
\subfloat{
\includegraphics[width=0.4\textwidth]{ 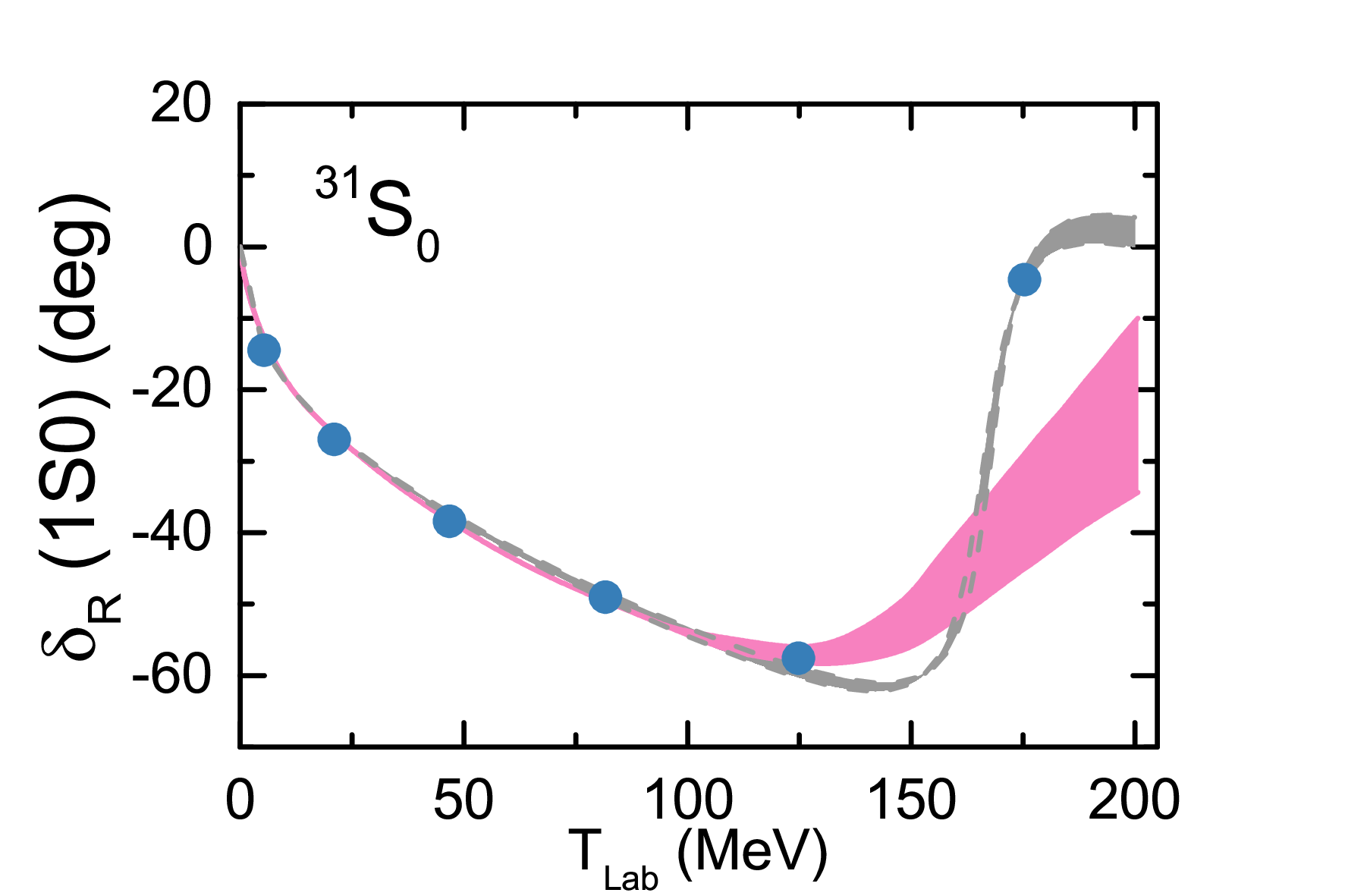}
}
\quad
\subfloat{
\includegraphics[width=0.4\textwidth]{ 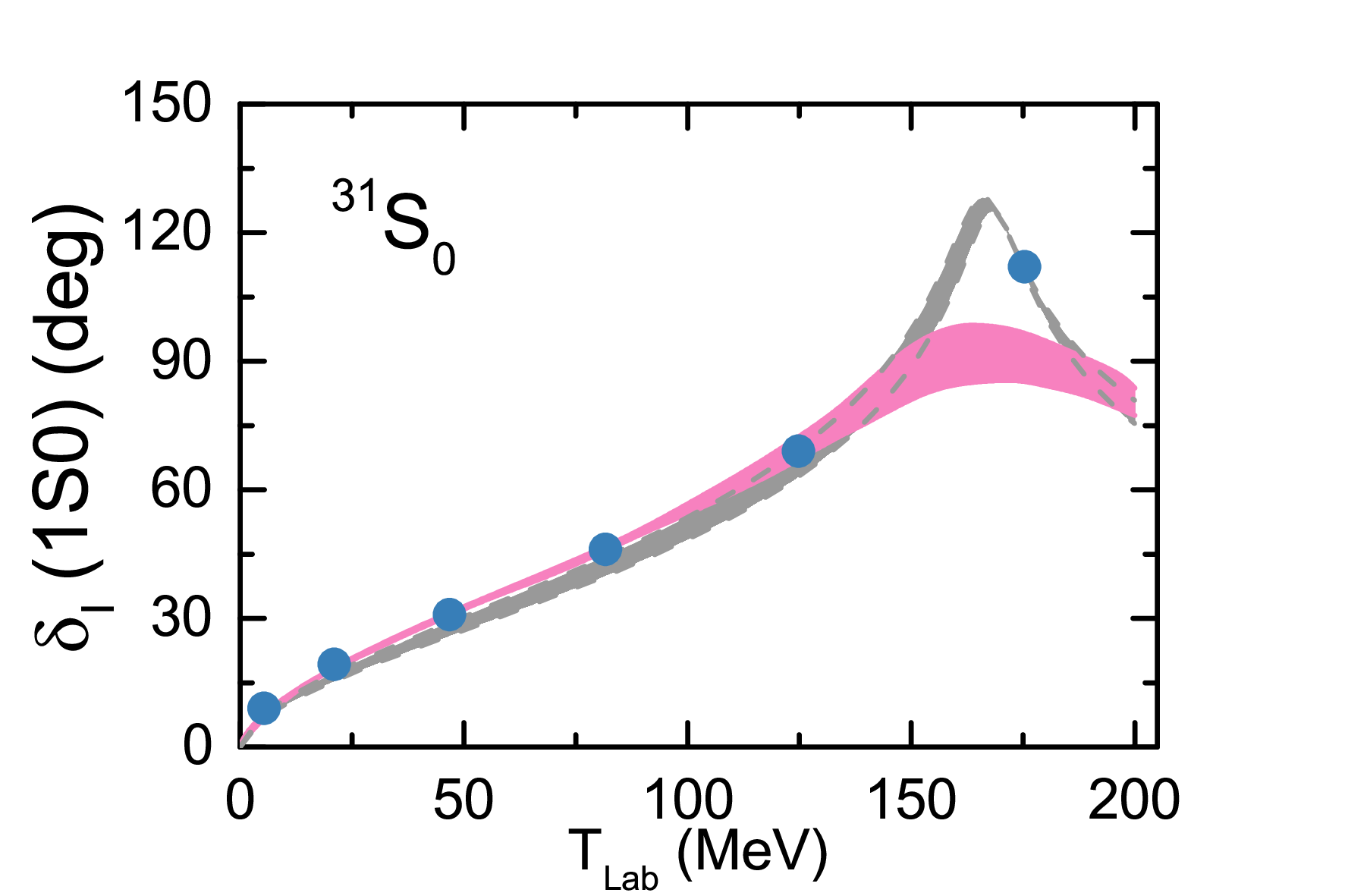}
}
\quad
\subfloat{
\includegraphics[width=0.4\textwidth]{ 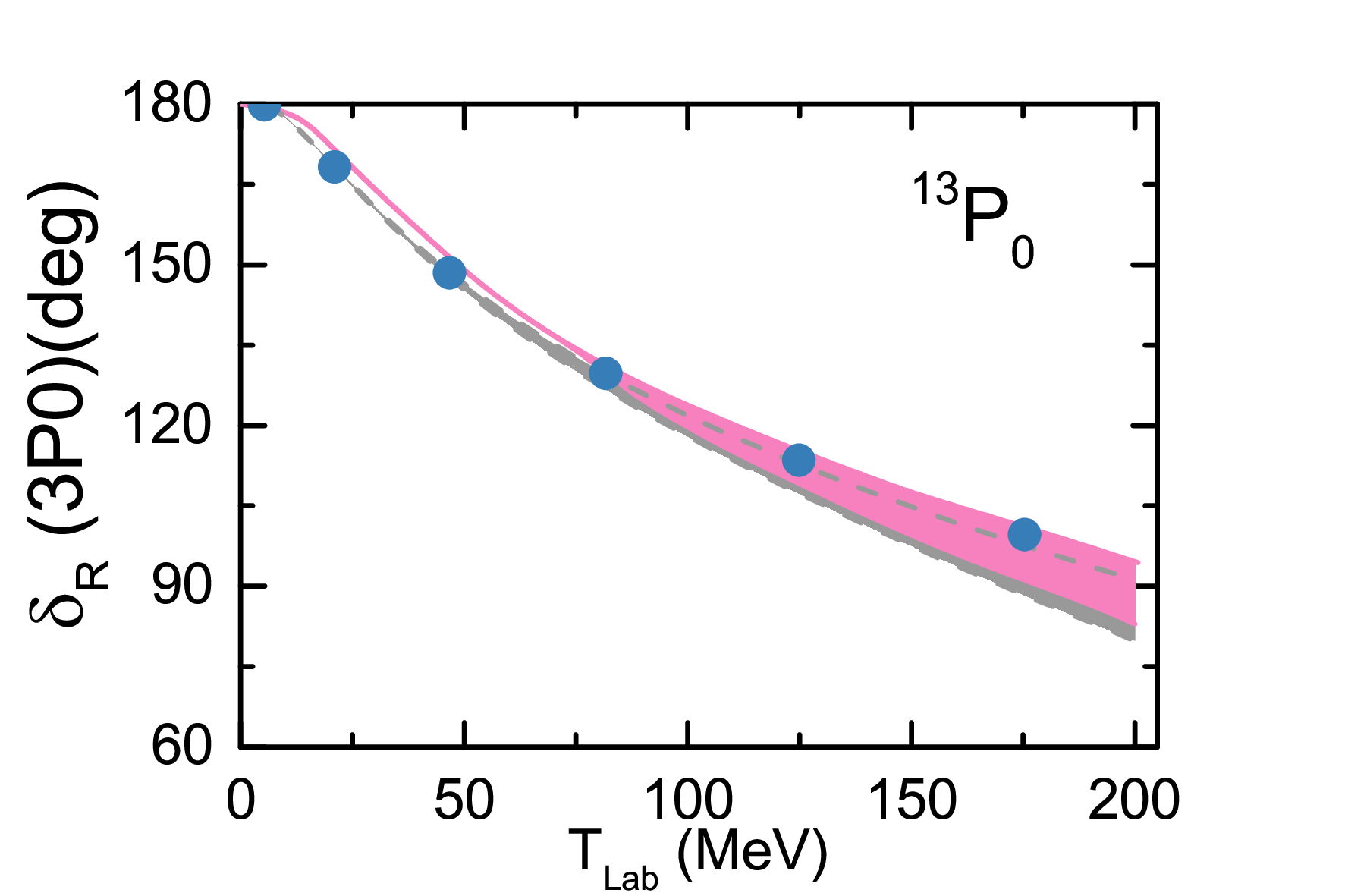}
}
\quad
\subfloat{
\includegraphics[width=0.4\textwidth]{ 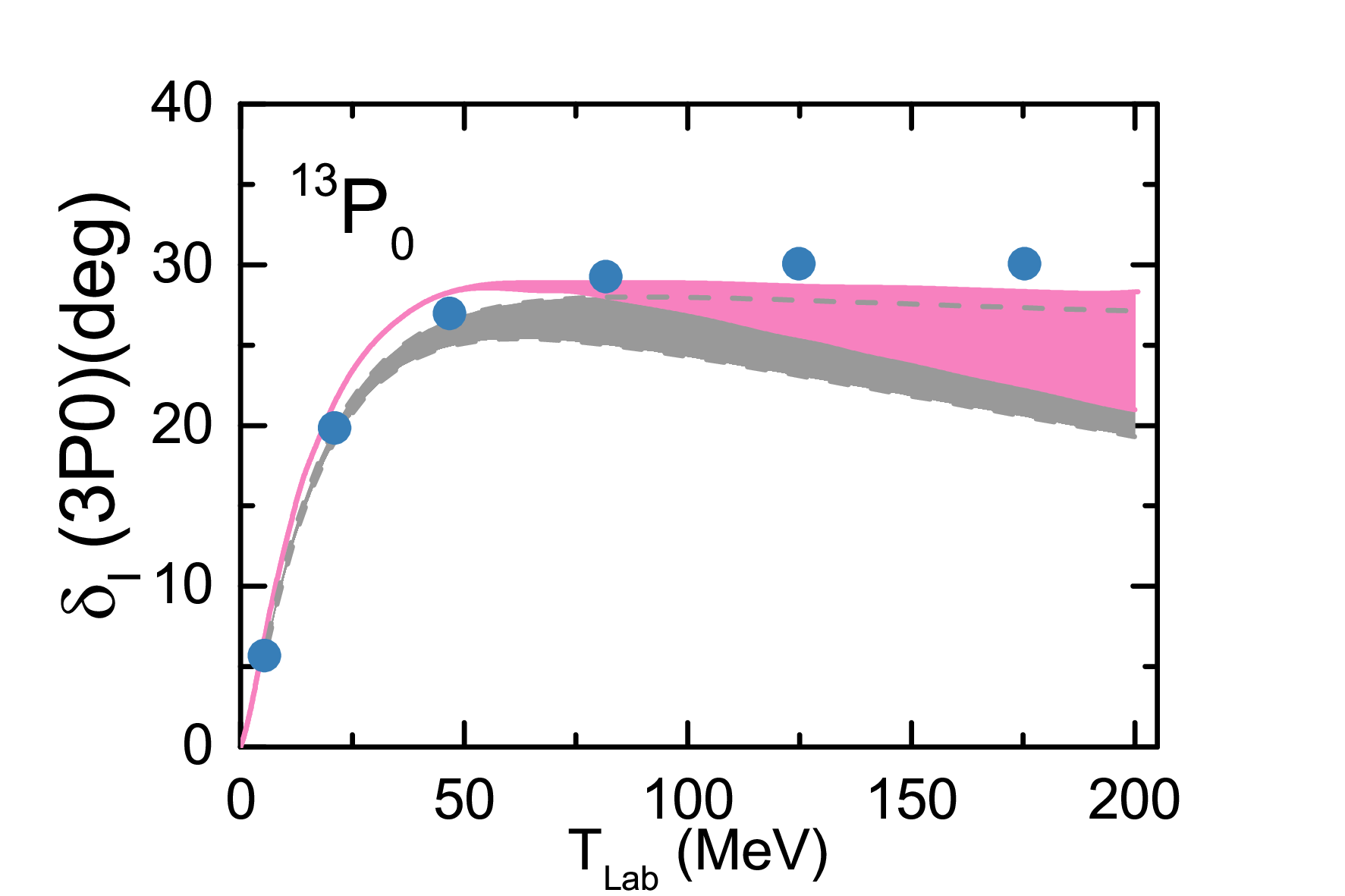}
}
\quad
\subfloat{
\includegraphics[width=0.4\textwidth]{ 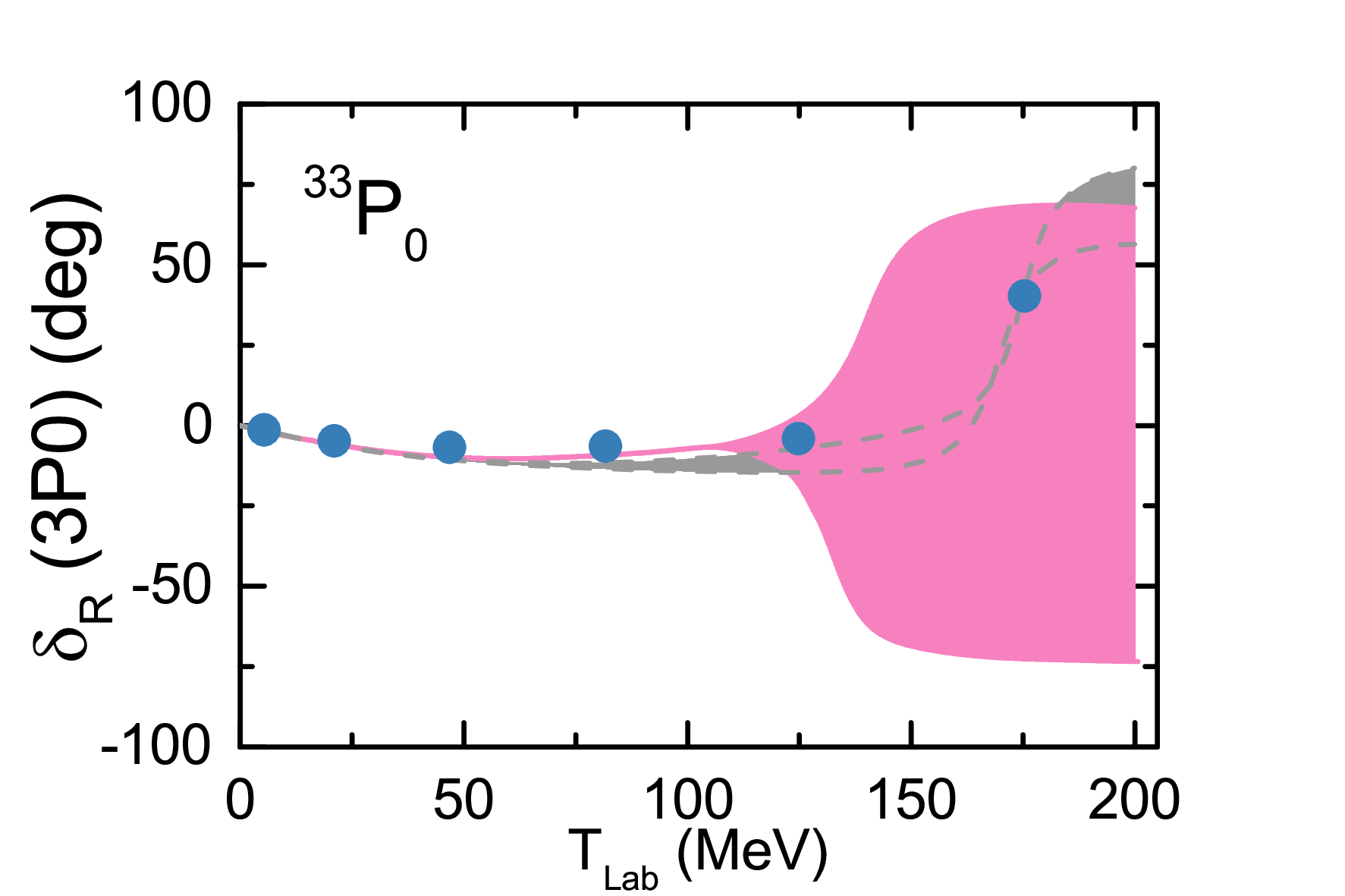}
}
\quad
\subfloat{
\includegraphics[width=0.4\textwidth]{ 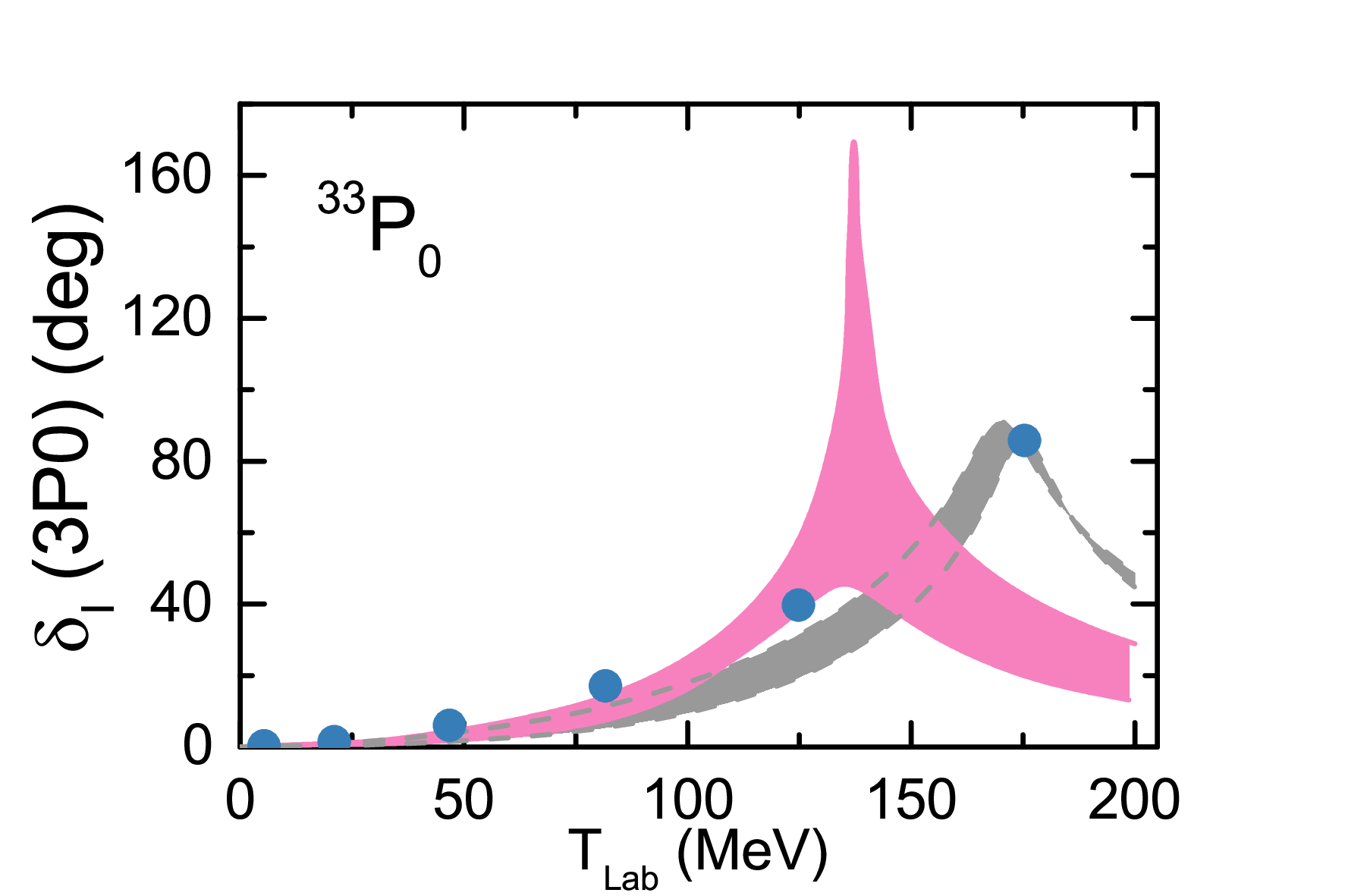}
}
\caption{Real and imaginary parts of the phase shift for the $^1S_0$ and $^3P_0$ partial waves. The gray bands show our results with the cutoff in the range
$\Lambda$ = 450–600 MeV. The pink bands show the NLO HB  chiral EFT results of Ref.~\cite{Kang:2013uia}. The blue dots refer to the solution of the PWA of Ref.~\cite{Zhou:2012ui}.}
\label{fig:1S03P0}
\end{figure*}

\begin{figure*}[htbp]
\centering
\subfloat{
\includegraphics[width=0.4\textwidth]{ 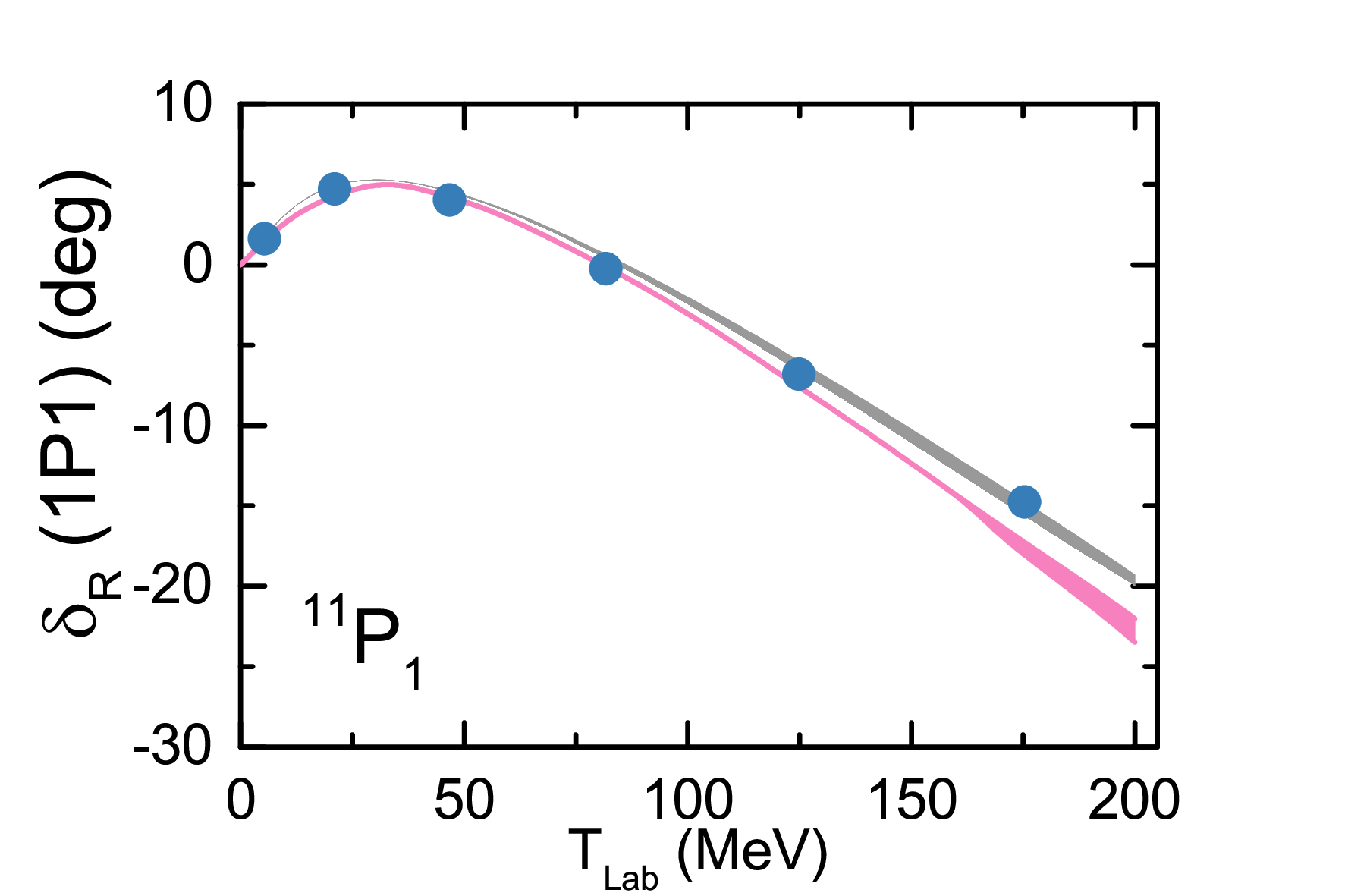}
}
\quad
\subfloat{
\includegraphics[width=0.4\textwidth]{ 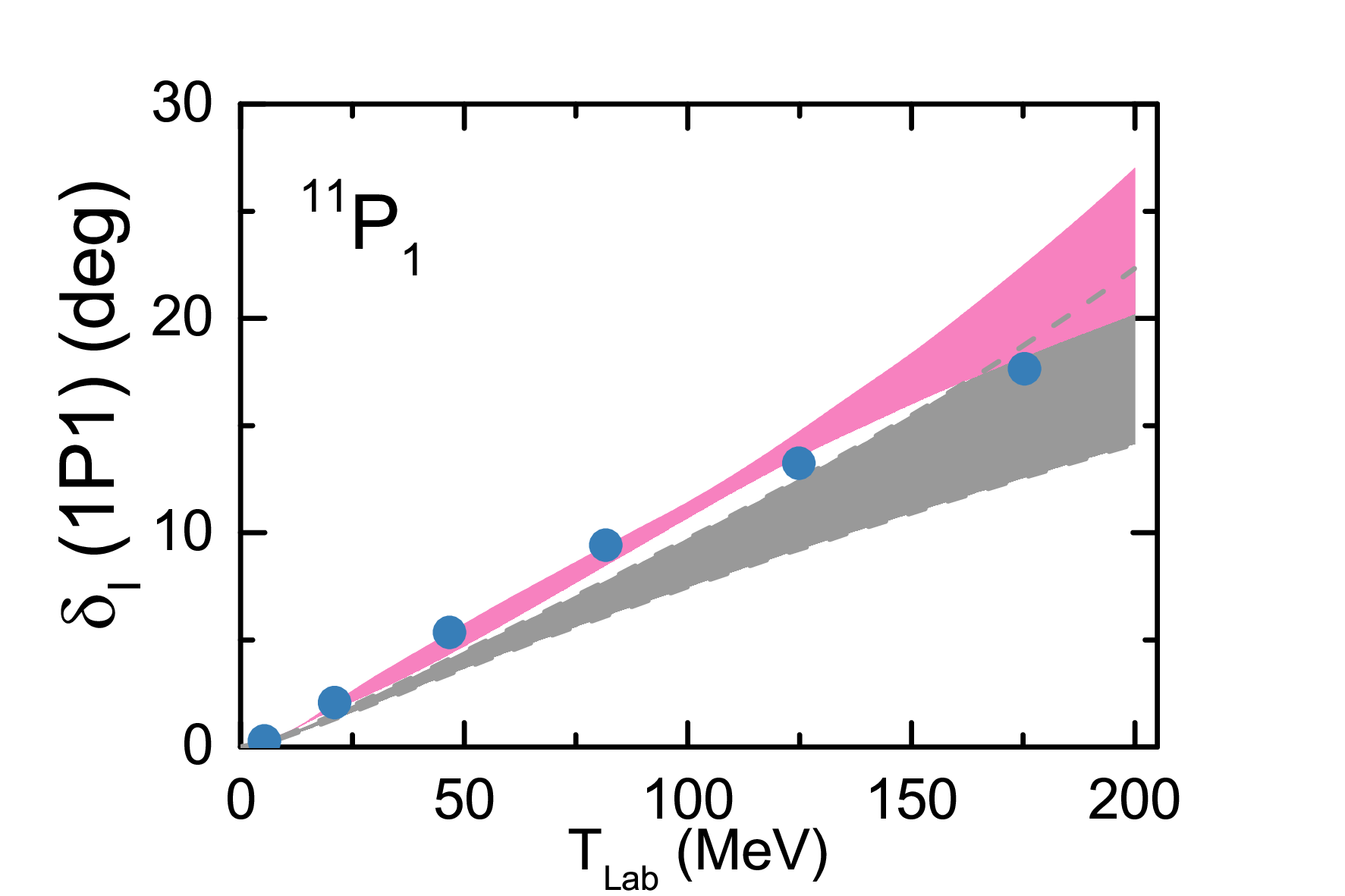}
}
\quad
\subfloat{
\includegraphics[width=0.4\textwidth]{ 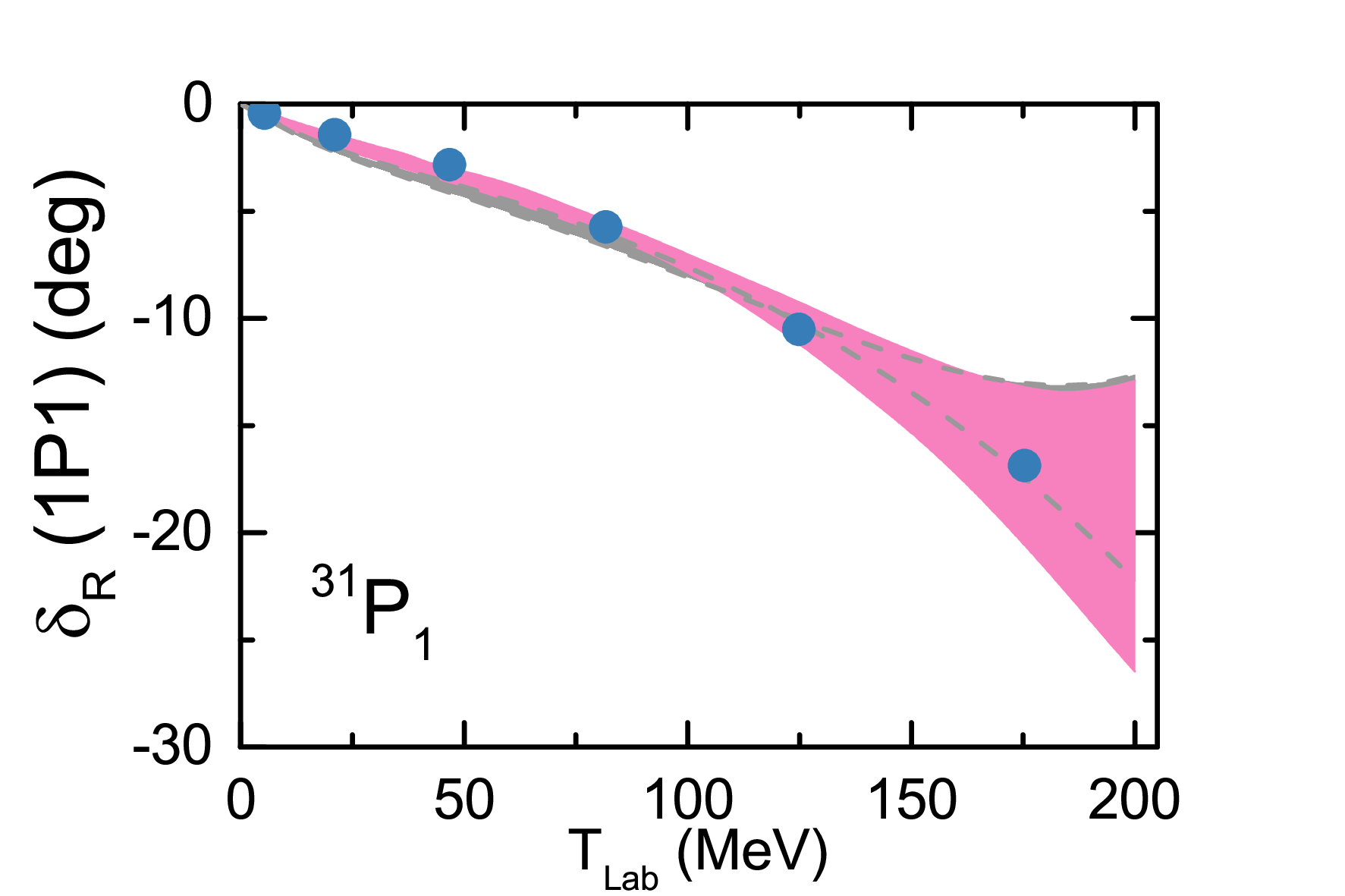}
}
\quad
\subfloat{
\includegraphics[width=0.4\textwidth]{ 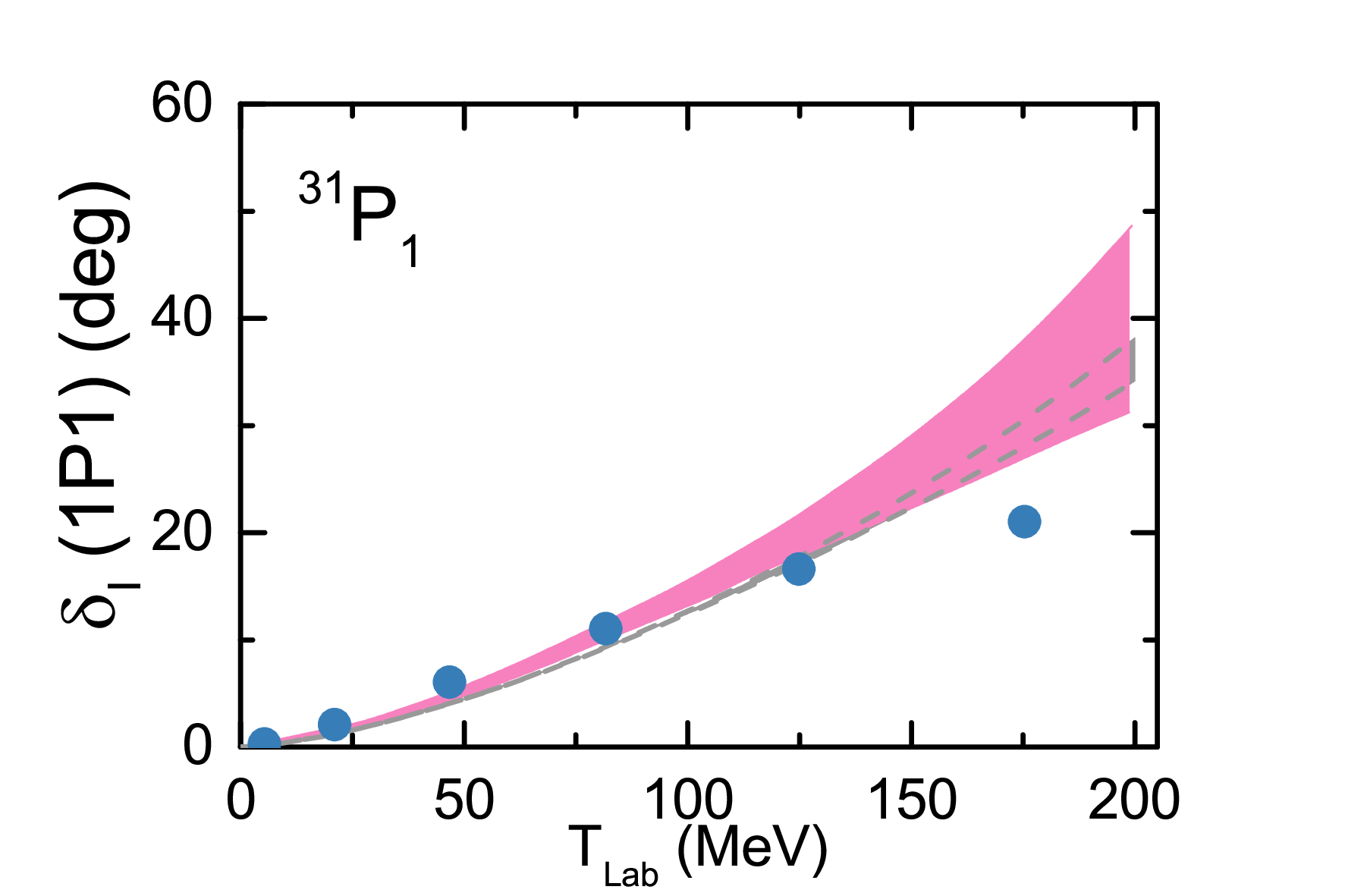}
}
\quad
\subfloat{
\includegraphics[width=0.4\textwidth]{ 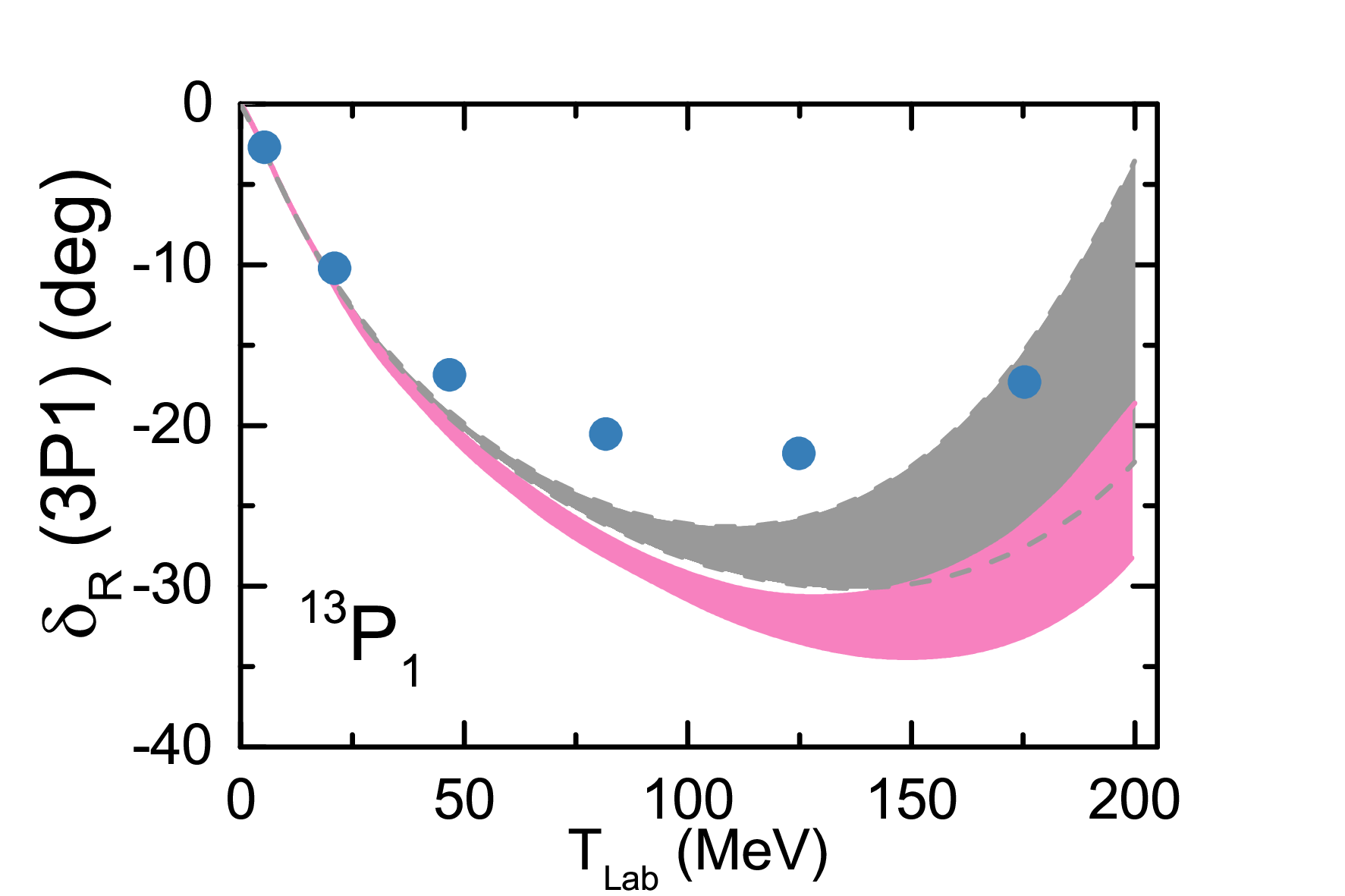}
}
\quad
\subfloat{
\includegraphics[width=0.4\textwidth]{ 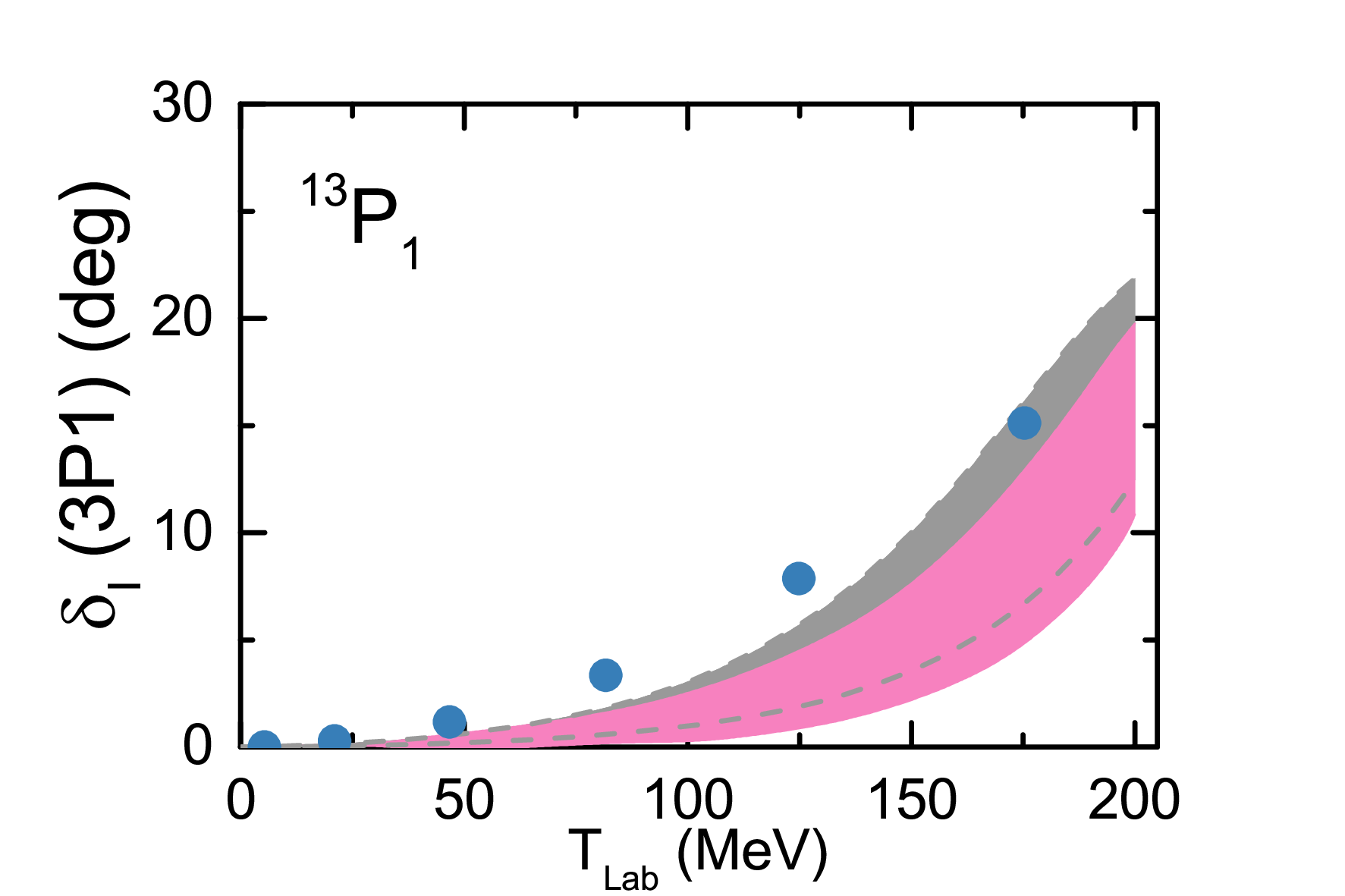}
}
\quad
\subfloat{
\includegraphics[width=0.4\textwidth]{ 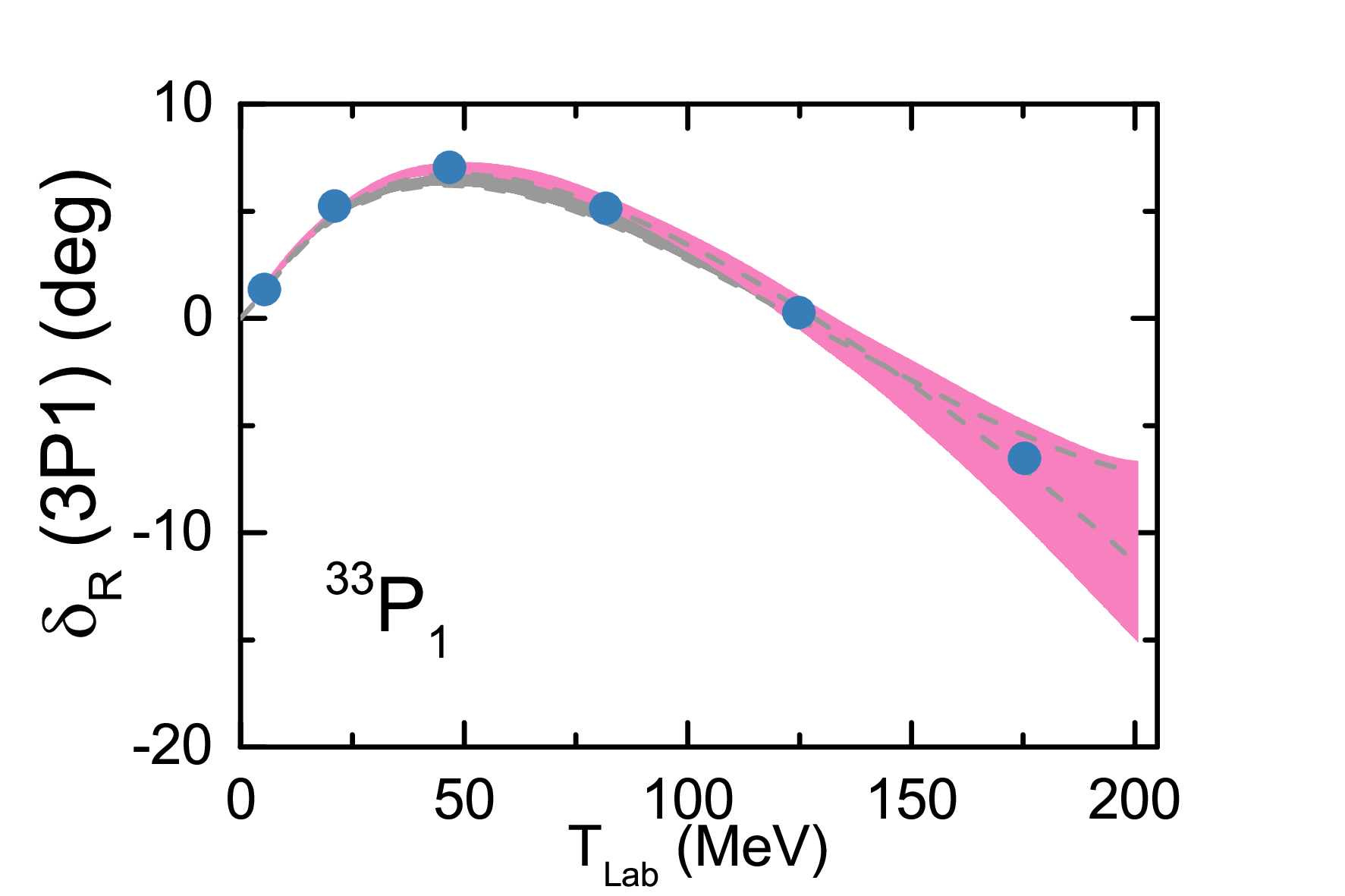}
}
\quad
\subfloat{
\includegraphics[width=0.4\textwidth]{ 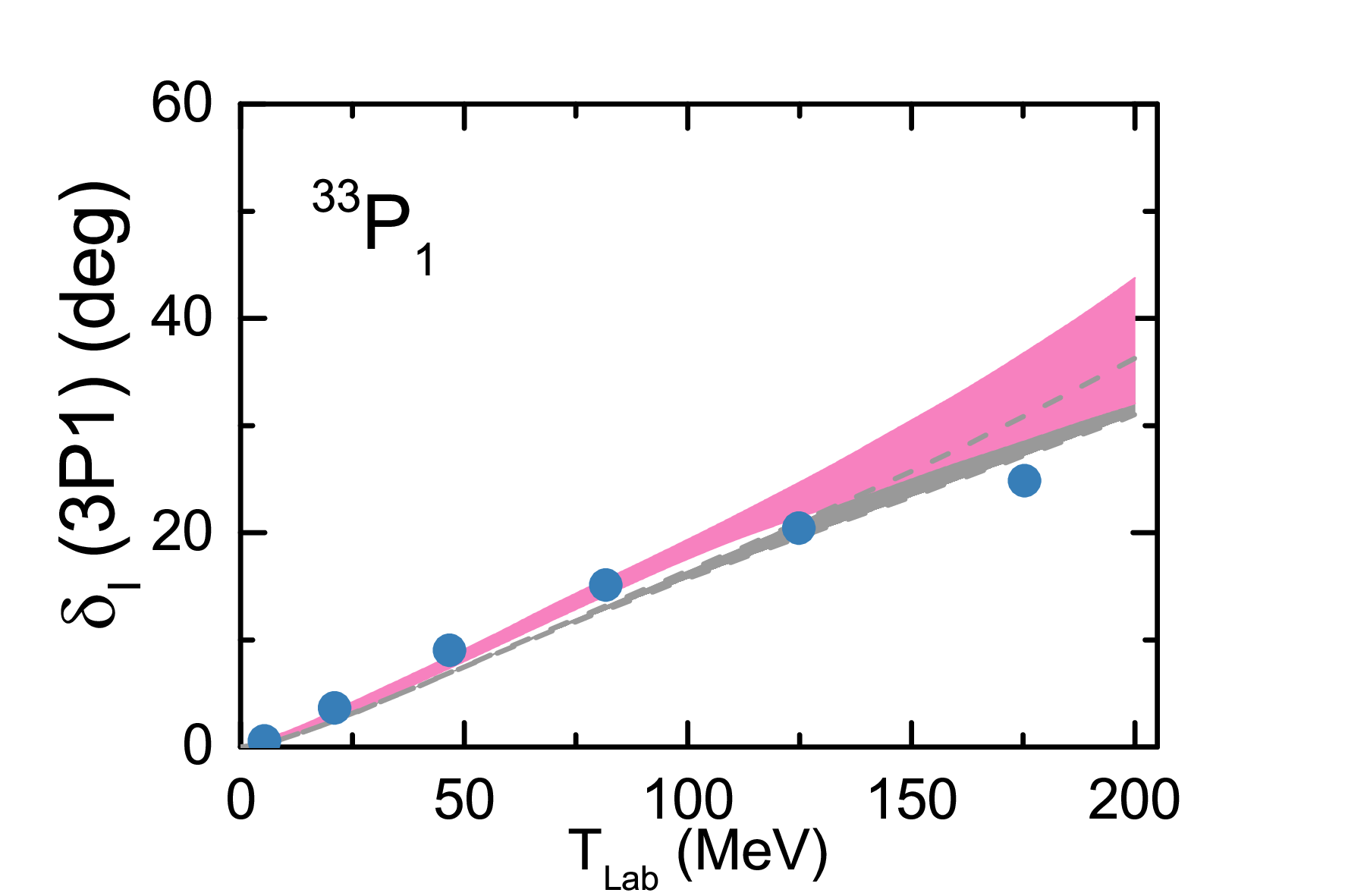}
}
\caption{Same as Fig.~\ref{fig:1S03P0}, but for the $^1P_1$ and $^3P_1$ partial waves.}
\label{fig:1P13P1}
\end{figure*}

\begin{figure*}[htbp]
\centering
\subfloat{
\includegraphics[width=0.4\textwidth]{ 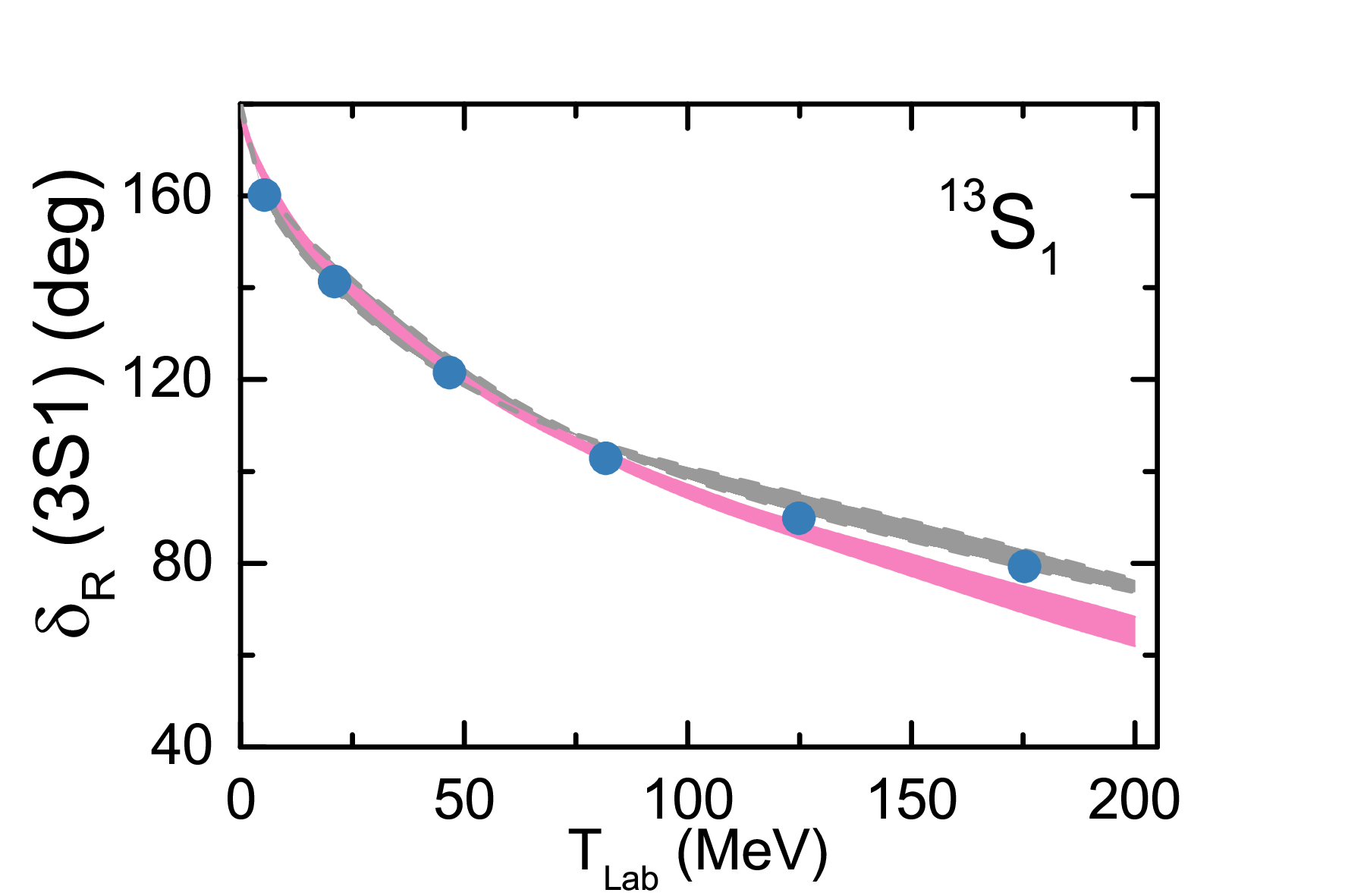}
}
\quad
\subfloat{
\includegraphics[width=0.4\textwidth]{ 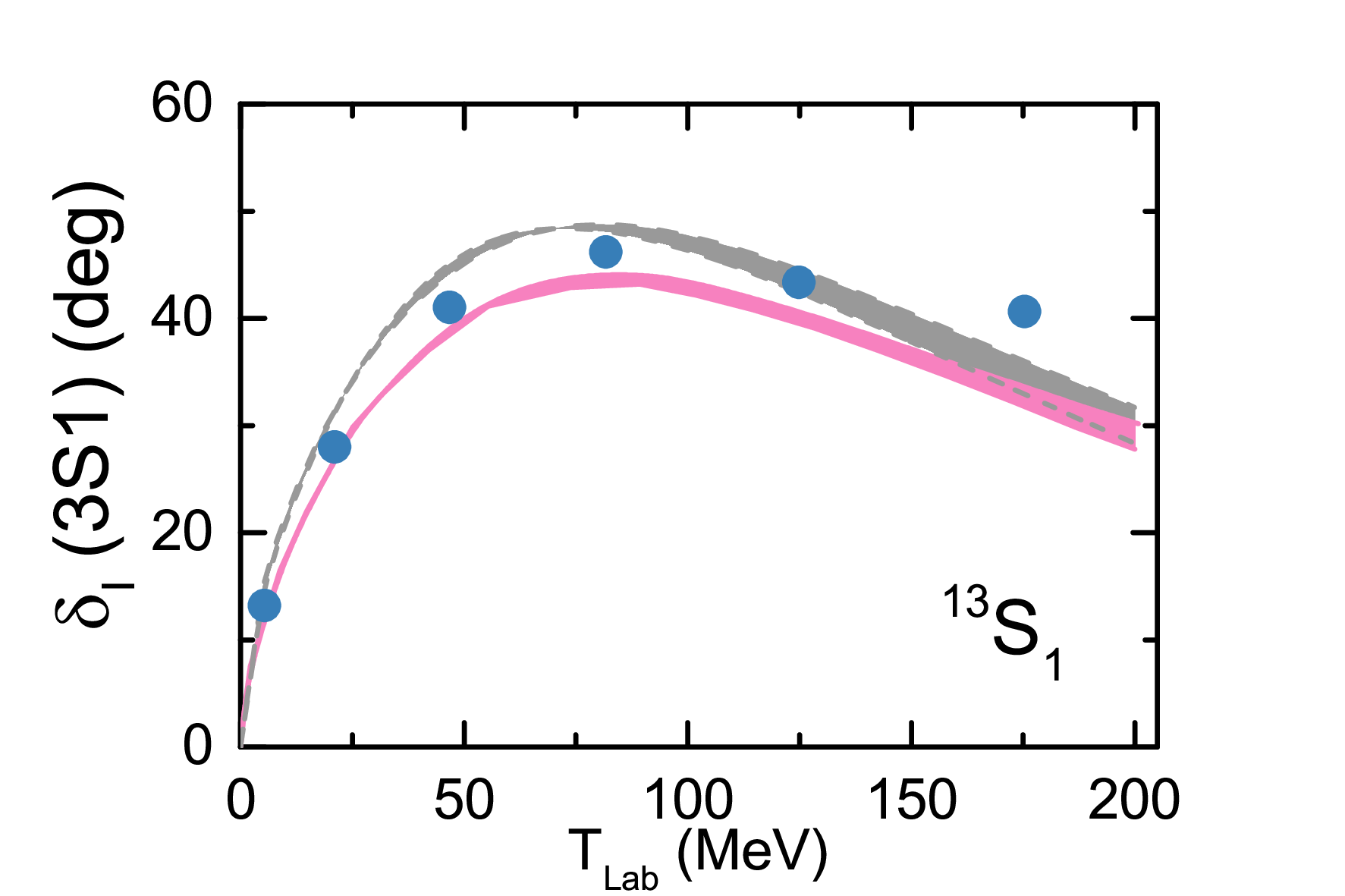}
}

\quad
\subfloat{
\includegraphics[width=0.4\textwidth]{ 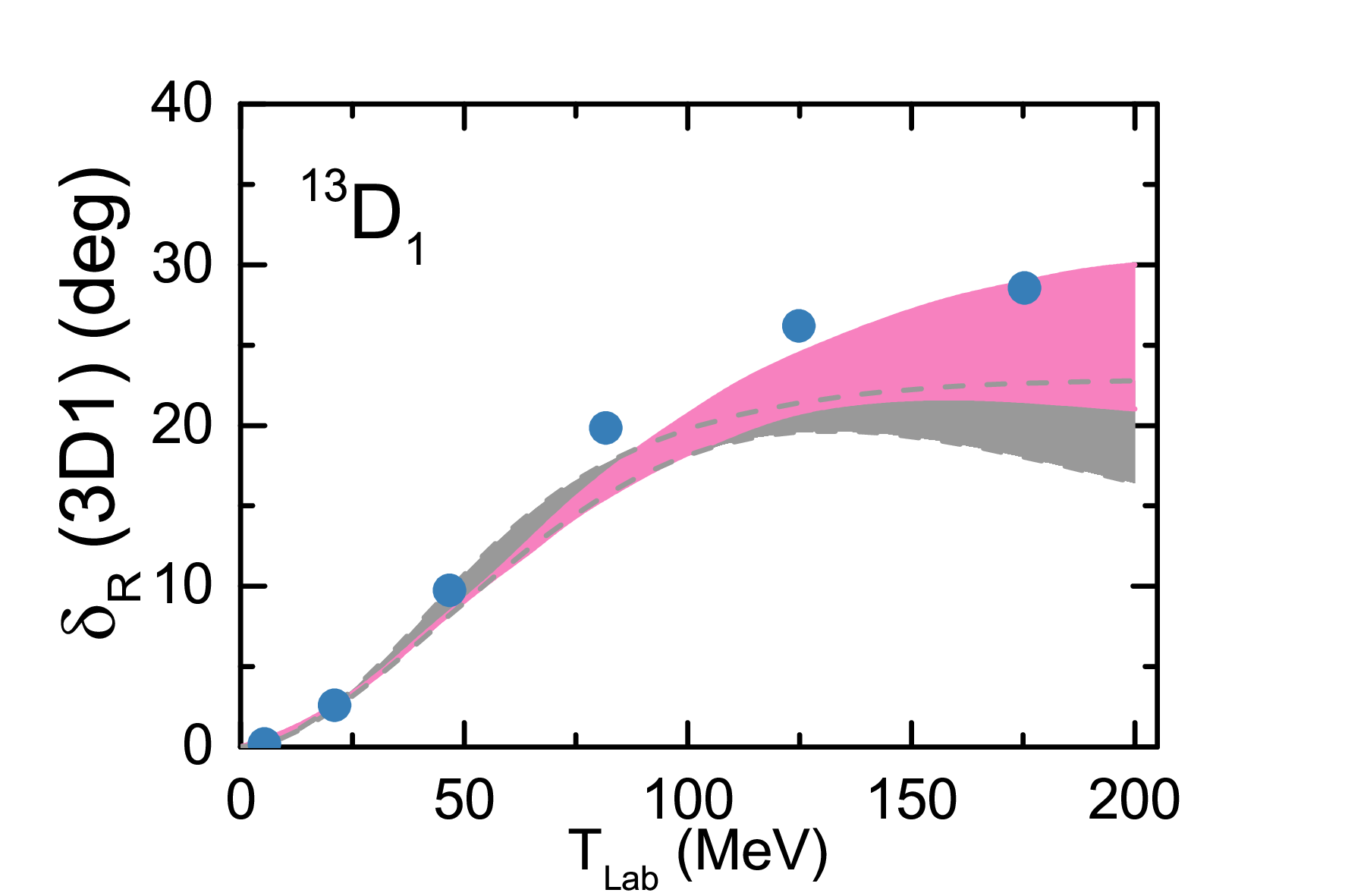}
}
\quad
\subfloat{
\includegraphics[width=0.4\textwidth]{ 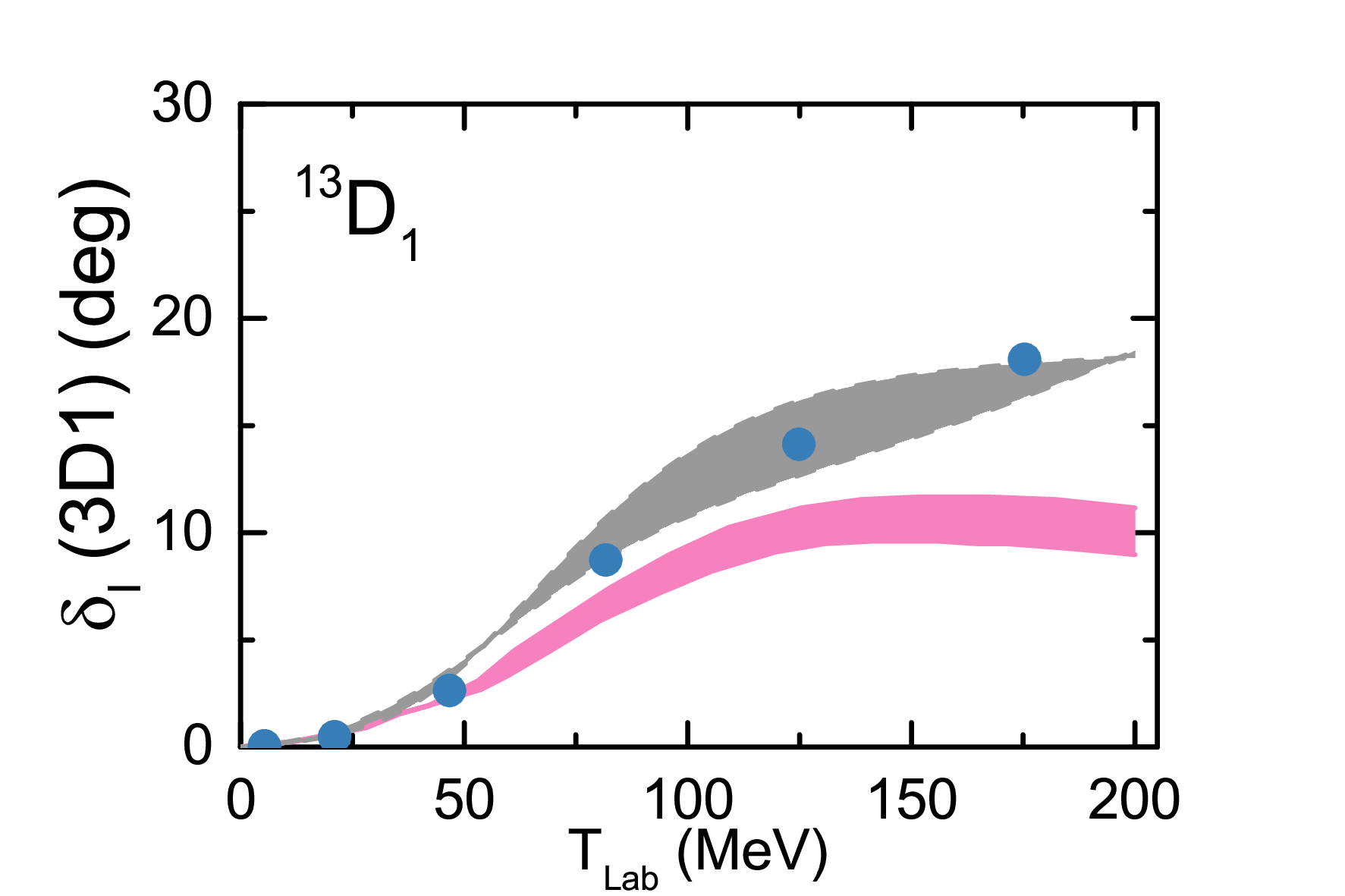}
}

\quad
\subfloat{
\includegraphics[width=0.4\textwidth]{ 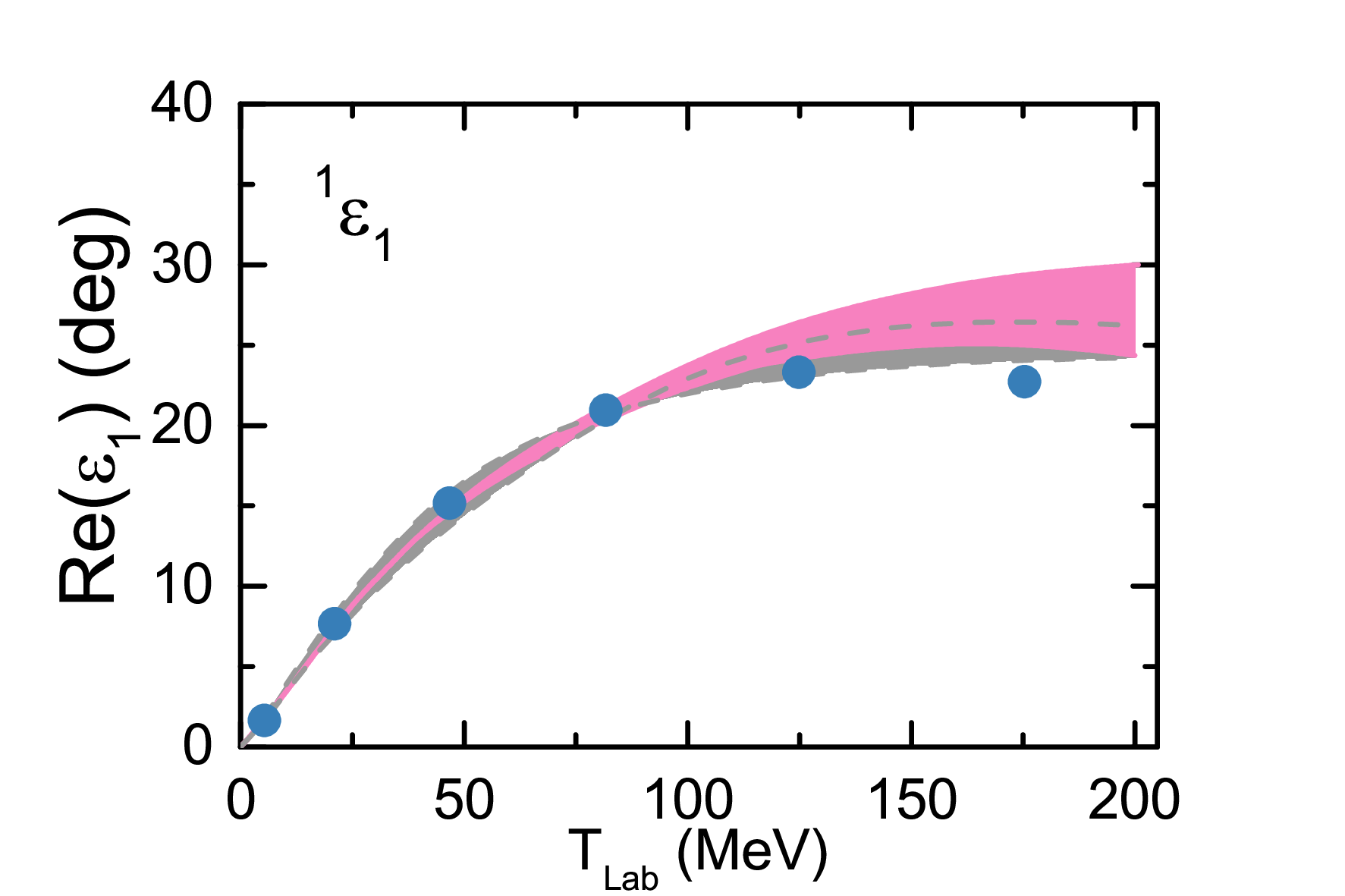}
}
\quad
\subfloat{
\includegraphics[width=0.4\textwidth]{ 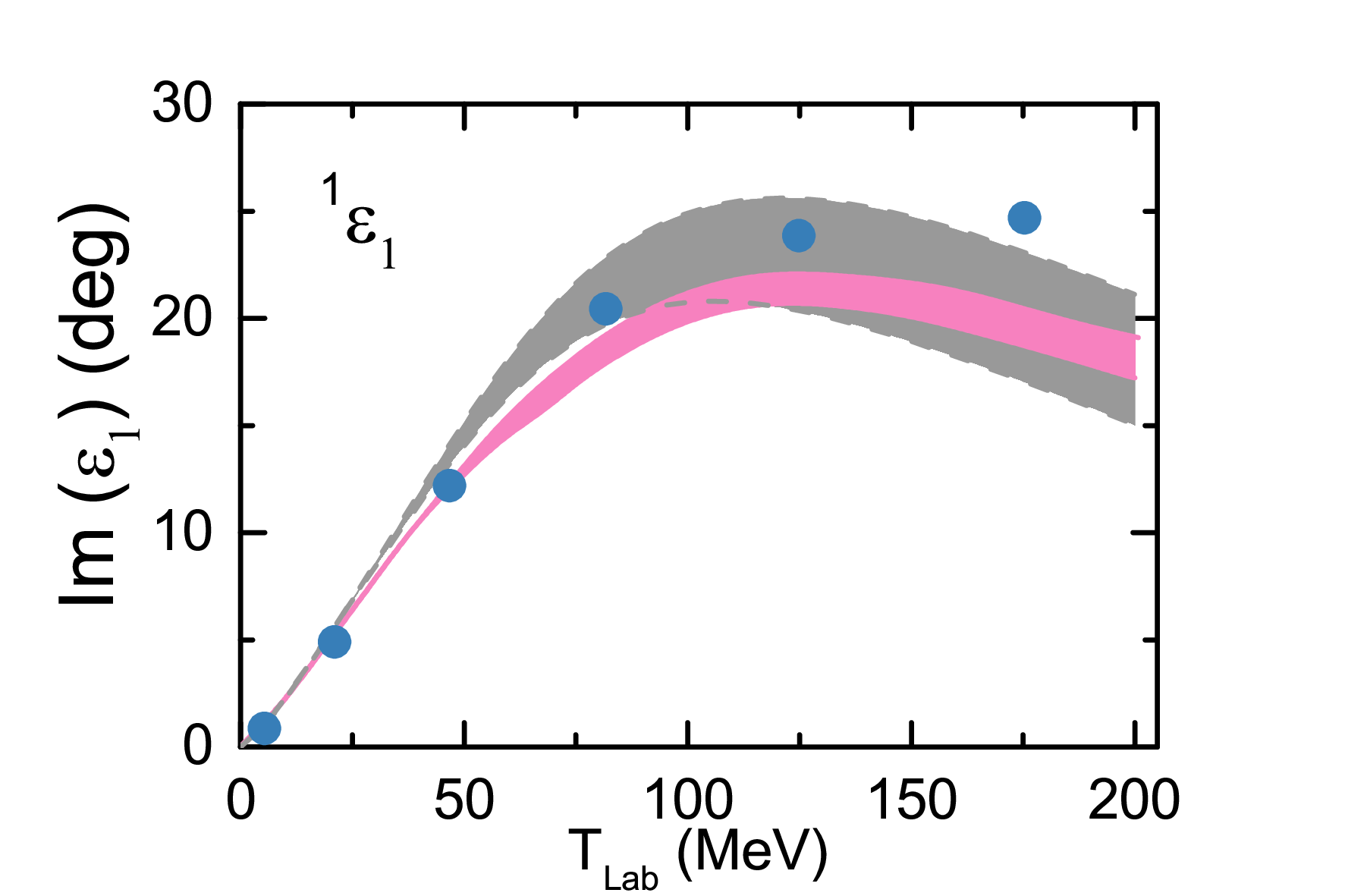}

}

\caption{Same as Fig.~\ref{fig:1S03P0}, but for the ${^3S_1} - {^3D_1}$ partial waves with $I=0$.}
\label{fig:13S113D1}
\end{figure*}

\begin{figure*}[htbp]
\centering
\subfloat{
\includegraphics[width=0.4\textwidth]{ 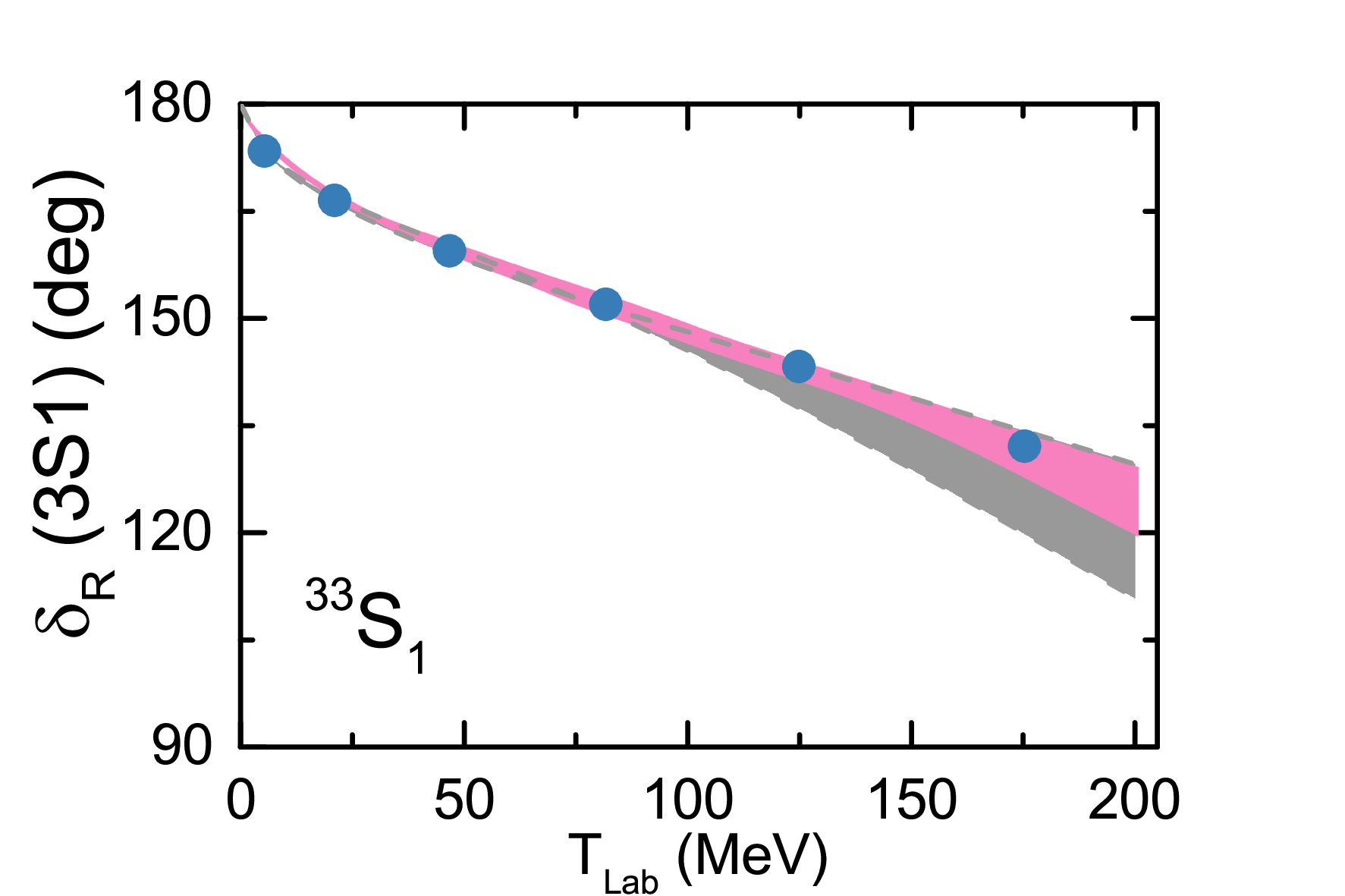}
}
\quad
\subfloat{
\includegraphics[width=0.4\textwidth]{ 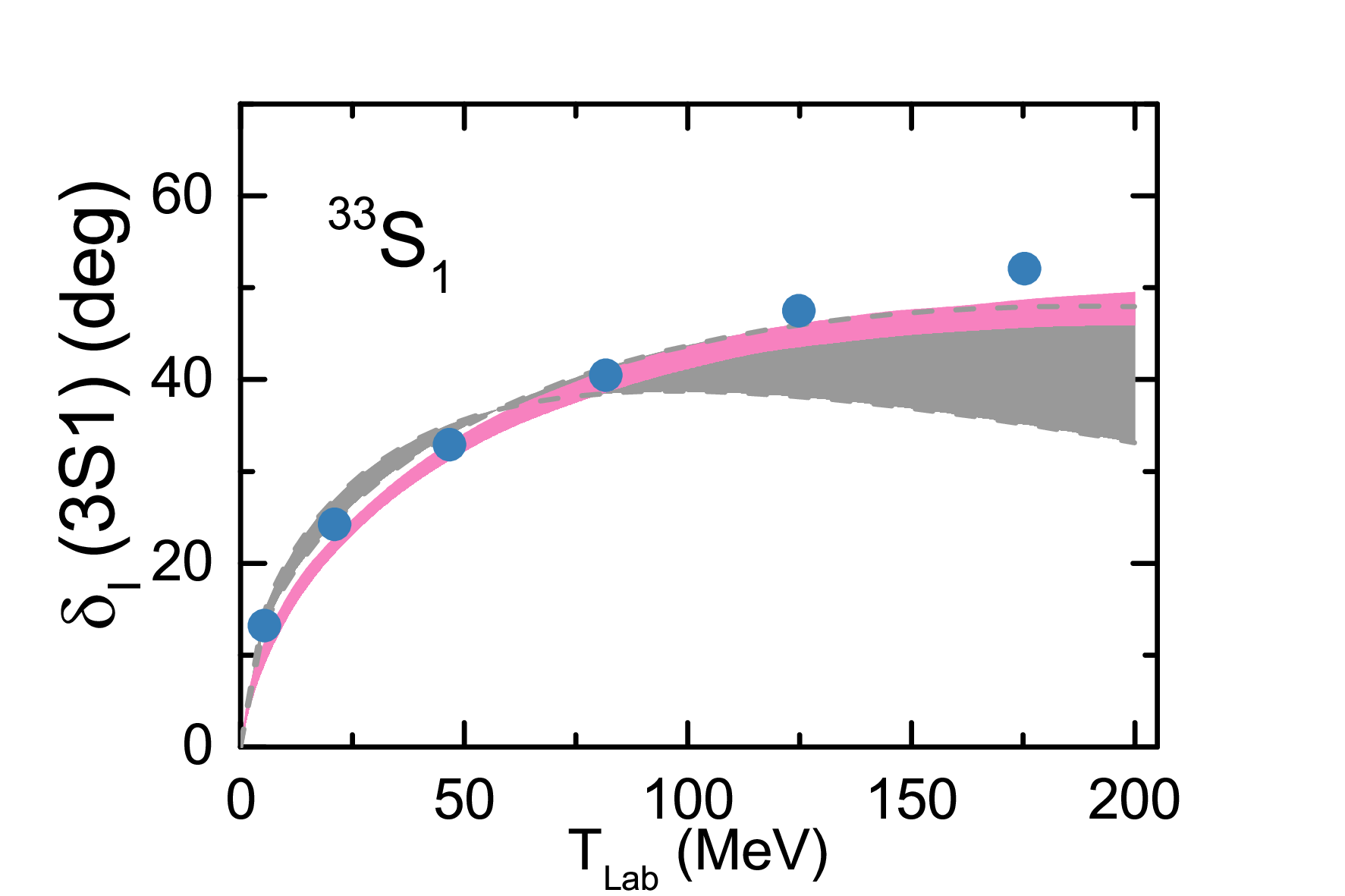}
}

\quad
\subfloat{
\includegraphics[width=0.4\textwidth]{ 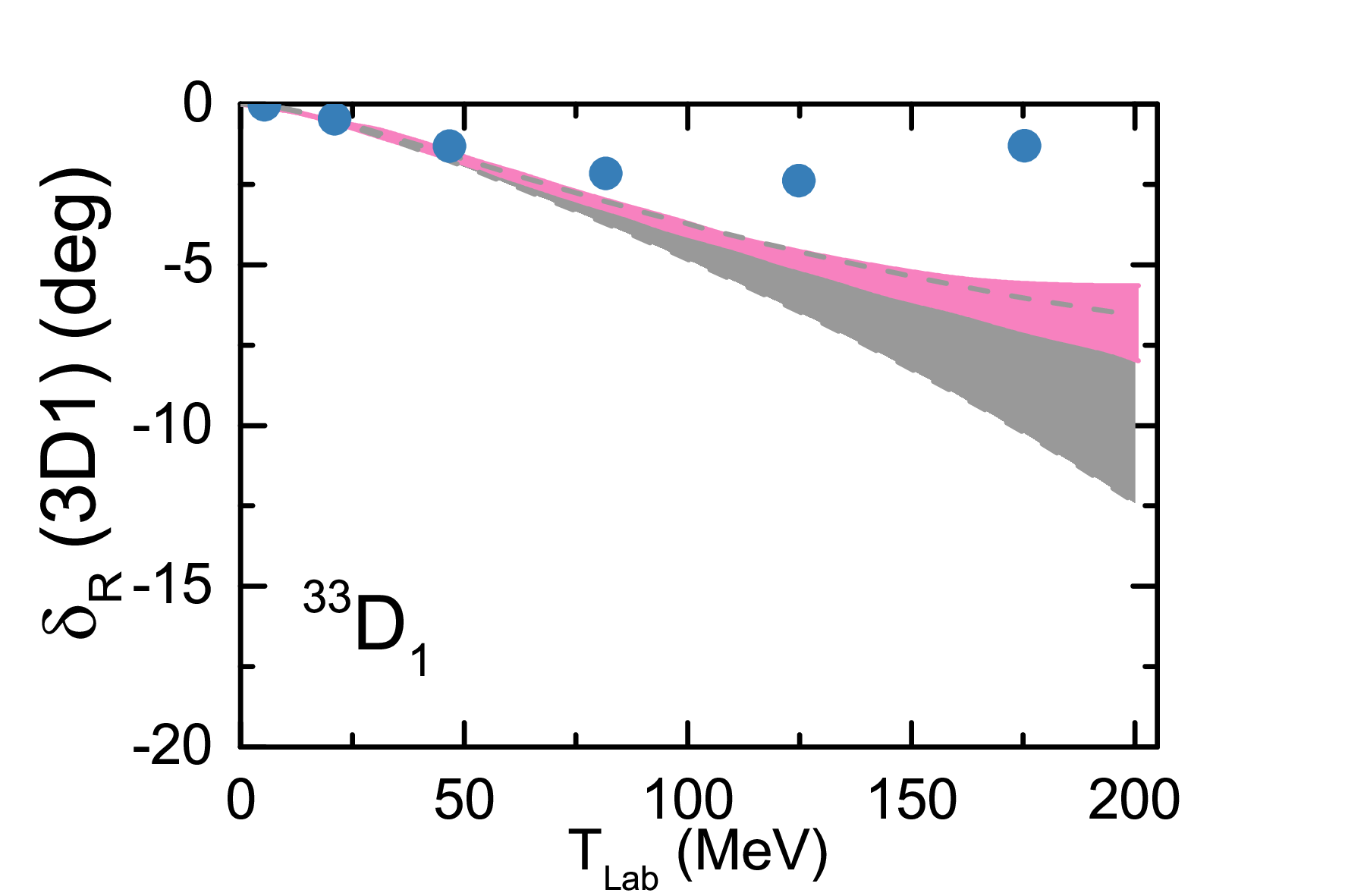}
}
\quad
\subfloat{
\includegraphics[width=0.4\textwidth]{ 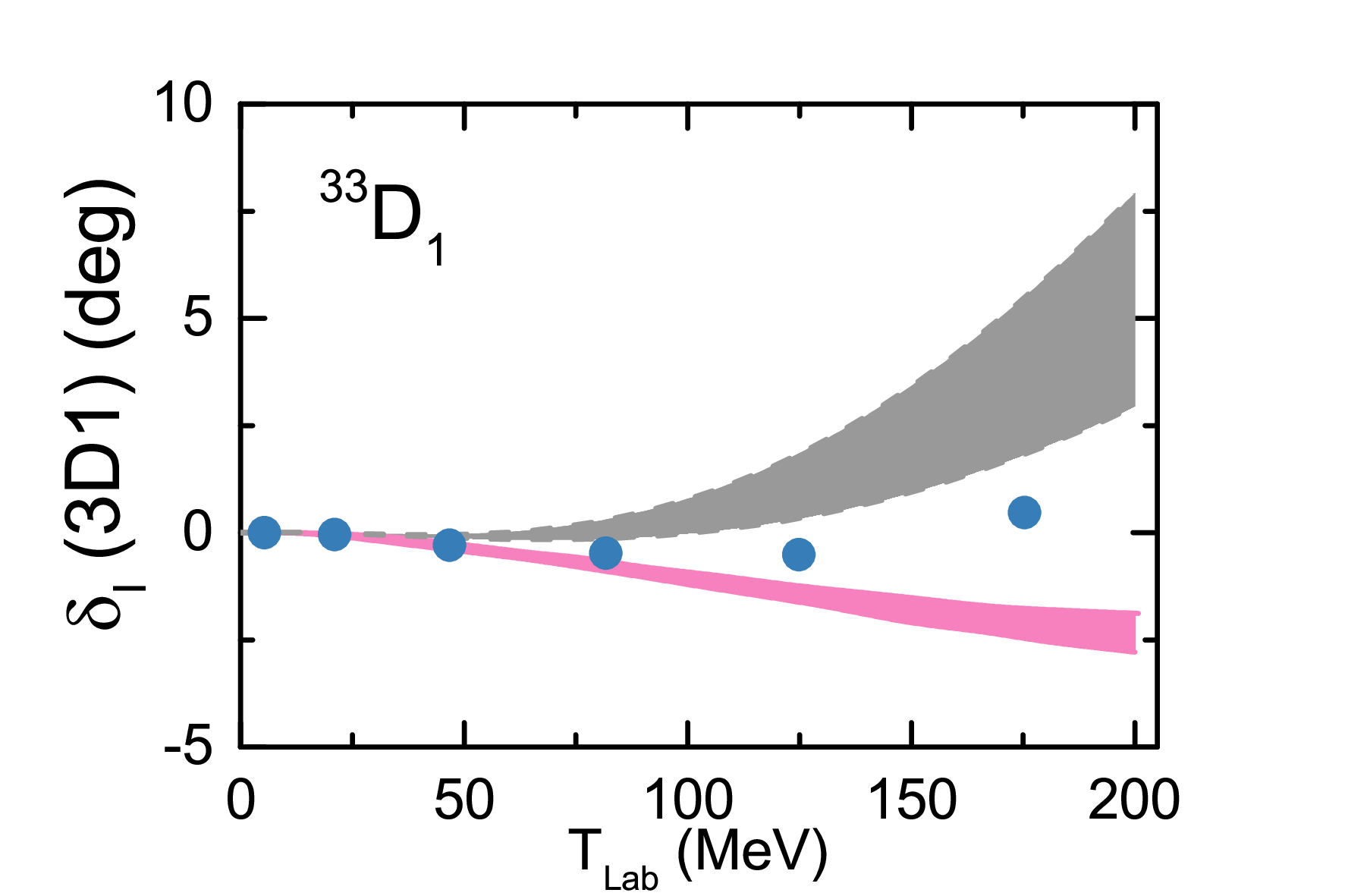}
}

\quad
\subfloat{
\includegraphics[width=0.4\textwidth]{ 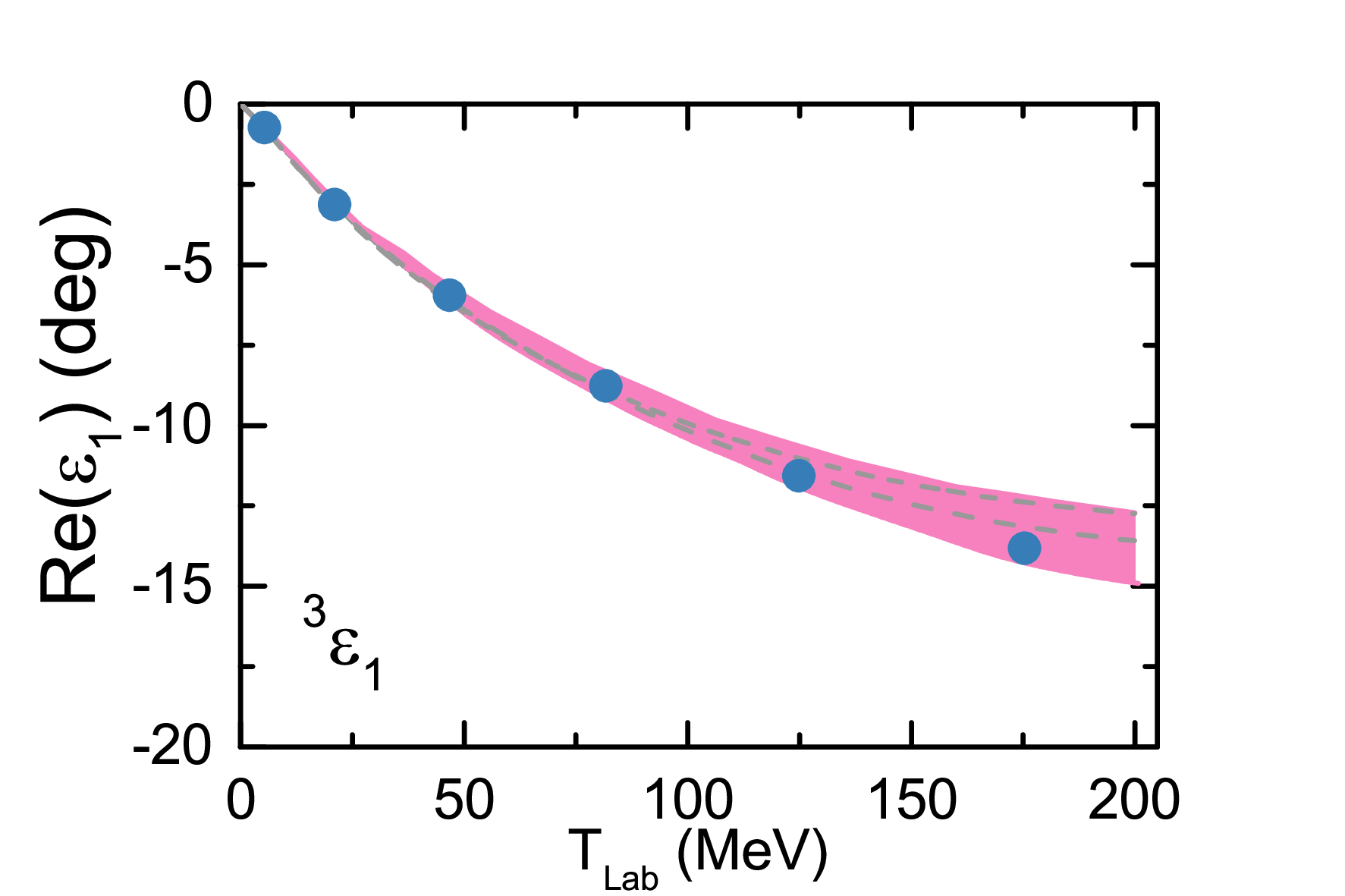}
}
\quad
\subfloat{
\includegraphics[width=0.4\textwidth]{ 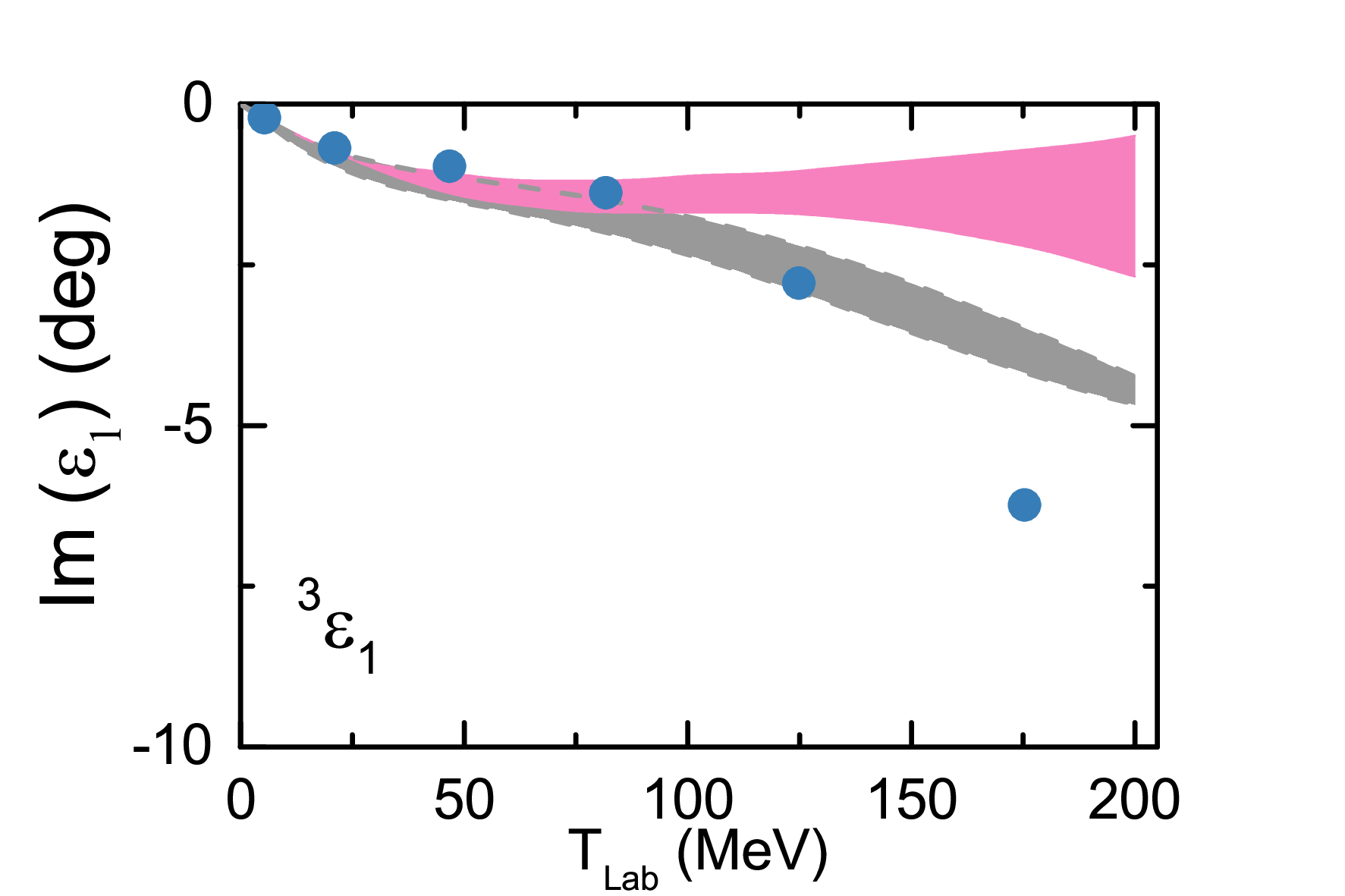}
}
\caption{Same as Fig.~\ref{fig:1S03P0}, but for the ${^3S_1} - {^3D_1}$ partial waves with $I=1$.}
\label{fig:33S133D1}
\end{figure*}

The phase shifts obtained in our study, the NLO HB results~\cite{Kang:2013uia}, and the $\bar{N}N$ PWA~\cite{Zhou:2012ui} for laboratory energies up to $200$ MeV are shown in Figs.~\ref{fig:1S03P0}-\ref{fig:33S133D1}. The partial waves are labeled in the spectral notation $^{\l(2I+1\r)\l(2S+1\r)}L_J$, and the bands are generated by varying the cutoff in the range $\Lambda=450-600$ MeV for both our hybrid method and the HB approach  The LO HB  phase shifts are not included for comparison because the potential relevant to the annihilation process, in this case, is only non-zero for the $^1S_0$ and $^3S_1$ partial waves. Hence, the descriptions of the phase shifts of other partial waves are very bad. In addition, even for the $^1S_0$ and $^3S_1$ partial waves, the differences between the LO HB  phase shifts and the PWA are significant compared with the differences between the NLO phase shifts and the PWA. 

Our results for the $J=0$ partial waves agree with the PWA for the energy region shown here. Compared with the NLO HB results, the overall cutoff dependence of the hybrid phase shifts is weaker, especially for the real parts of the phase shifts of the $I=1$ partial waves. At the same time, one can observe a sizeable cutoff dependence in the NLO HB results for
energies above $150$ MeV because of the resonance-like behaviors. Since the number of free parameters for the $J=0$ partial waves in the hybrid method and the NLO HB formalism is identical ($4$ for the $^1S_0$ partial wave and $2$ for the $^3P_0$ partial wave, including annihilation parameters), the relatively weaker cutoff dependence has to do with the relativistic corrections of the scattering equation and the potentials at orders higher than $\mathcal{O}\l(p^2\r)$ (in the conventional Weinberg power counting). Note that the NLO HB  results are obtained by fitting to the PWA of Ref.~\cite{Zhou:2012ui} at $p_{\text{lab}}\leq 500$ MeV, while in our study one more datum at $p_{\text{lab}}=600$ MeV is also included in the fitting process for the $^{31}S_0$ and $^{33}P_0$ partial waves as explained above. However, adapting the same fitting strategy in the NLO HB  framework ruins the descriptions of PWA at $p_{\text{lab}}\leq 500$ MeV. Therefore, the improvement in the cutoff dependence in our  framework cannot be completely attributed to the differences in the fitting procedures. An exception exists in the imaginary part of the phase shift of the $^{11}S_0$ partial wave, where the cutoff dependence of the hybrid  results is sizeable at laboratory energies above $150$ MeV. This is related to the description of the $^{13}P_1$ partial wave, whose PWA yields a negative phase at low energies, which tends to become positive at higher energies. As argued in Ref.~\cite{Kang:2013uia}, reproducing such phase shifts requires a repulsive potential at large separations of the antinucleon and nucleon but becomes attractive at short distances. Since the real parts of the potential for $^{3}P_1$ are controlled by the LECs in the $^1S_0$ partial wave as shown in Eq.~\eqref{eq:LO_CT_LSJ}, the description of $\delta_\text{I} \l( ^{11}S_0 \r) $ is influenced by the demand for such an attractive potential.  Improvement might be possible by using an NLO covariant potential for the elastic process. 

For the $J=1$ uncoupled channels, the hybrid  and NLO HB  results are comparable. The NLO HB  results are better for the imaginary part of the phase shift in the $^{11}P_1$ partial wave. In comparison, the hybrid  results are better for the real part of the phase shift in the $^{31}P_1$ partial wave. Still, the relativistic corrections are not attractive enough to account for the discrepancies between the calculated phase shifts and the PWA. As for other partial waves, both results are comparable, but the cutoff dependence of the hybrid  results is weaker than that of the NLO HB  results at laboratory energies above 150 MeV, in analogy to the results for the $^{31}S_0$ and $^{33}P_0$ partial waves. It should be emphasized that the real parts of the hybrid  potentials for the $^{1}P_1$ and $^{3}P_1$ partial waves are determined by the $S$- wave LECs as shown in Eq.~\eqref{eq:LO_CT_LSJ}. In contrast, the real parts of the NLO HB potentials contain as many LECs as the imaginary parts. Thus, the improvements in the cutoff dependence must originate from the relativistic corrections. 

For the $J=1$ coupled channels, the $S$-wave phase shifts are generally well reproduced. The  $^3D_1$ phase shift and mixing angle $\epsilon_1$ show strong cutoff dependence. However, it is not so surprising since they have no free parameters. The intriguing thing is that the relativistic corrections shift the trends of $\delta_\text{I} \l(^{33}D_1 \r)$ and $\text{Im} \l( \epsilon_1 \r)$ to the right direction at laboratory energies above $100$ MeV compared to the NLO HB results. However, the correction seems too large for the $^{33}D_1$ partial wave. As a result, the cutoff dependence of $\delta_\text{I} \l( ^{33}S_1\r)$ becomes large at that energy region.

We want to emphasize that although the interaction constructed in this work is not fully covariant in the sense that the contributions of all the annihilation channels $V_{\bar{N}N \rightarrow X}$  are expanded in powers of antinucleon/nucleon three momenta, the present method is reasonable at least at a qualitative level according to the previous studies in the
$NN$ sector~\cite{Ren:2016jna}. Considering that the contributions irrelevant to the annihilation process are calculated fully covariantly at LO, we think the above discussions on ``power counting" and ``convergence" are reasonable.

Next, we turn to the near-threshold $\bar{N}N$ structures. The phase shifts shown in Figs.~\ref{fig:1S03P0}-~\ref{fig:33S133D1} suggest the existence of bound states in the $^{11}S_0, ^{13}P_0, ^{13}S_1,$ and $^{33}S_1$ channels because their phase shifts are about $180^{\circ}$ at threshold. Therefore, we search for possible $\bar{N}N$ bound states in these channels. The corresponding binding energies obtained with our hybrid potential and the NLO HB  potential~\cite{Kang:2013uia} are summarized in Table~\ref{tb:bound_states}. Although these structures have complex $E_B$, and the sign of the real part
of $E_B$ is even positive in some cases, according to Refs.~\cite{Kang:2013uia, Badalian:1981xj}, the poles that we found can still be referred to as bound states because they lie on the
physical sheet and move below the threshold when the annihilation potential is switched off.  Moreover, we  find a deeply bound state with $E_B = \l( -102.2 ,-152.5 \r) - \text{i} \l( 79.1 , 199.3 \r)$ MeV in the $^{11}S_0$ channel, whose quantum number is consistent with the pseudoscalar interpretation of $X(1835)$, $X(1840)$, and $X(1880)$ suggested by the BESIII Collaboration~\cite{BES:2005ega,BESIII:2023vvr}, despite that it is located far below the $\bar{N}N$ threshold and our result suffer relatively large uncertainties. A firm conclusion can only be drawn once reliable theoretical uncertainties can be estimated. We want to mention that the studies employing the semi-phenomenological $\bar{N}N$ interactions have found a bound state in the $^{11}S_0$ channel~\cite{Yan:2004xs, Sibirtsev:2004id,Ding:2005ew, Dedonder:2009bk}, although the predicted binding energies are rather different. Therefore, more studies are  needed to confirm the nature of this state. Apart from the bound states, the phase shifts exhibit resonance-like structures in $^{31}S_0$ and $^{33}P_0$ partial waves at energies above $150$ MeV. Thus, we also look for poles in the second Riemann sheet in these two channels. However, we do not find any resonant states in this energy region.

\begin{table}[h!]
\centering
\caption{$\bar{N}N$ bound states and their binding energies. The uncertainties originate from the variation of the cutoff in the range $\Lambda=450-600$ MeV.}
\label{tb:bound_states}
\resizebox{\linewidth}{!}{
\begin{tabular}{c|c|c}
\hline
\hline
  \multirow{2}{*}{Partial Wave}   & \multicolumn{2}{c}{$E_B$ (MeV)} \\
  \cline{2-3}
  & Hybrid  & NLO HB ~\cite{Kang:2013uia}\\
     \hline
    $^{11}S_0$ & $\l( -102.2 ,-152.5 \r) - \text{i} \l( 79.1 , 199.3 \r)$ & /~\footnote{\label{note1}It is unclear whether a $\bar{N}N$ bound states can be observed in this channel because the possible structures are only searched for near $\bar{N}N$ threshold. }\\ 
    $^{13}P_0$ & $\l( -1.5 , -2.1 \r) - \text{i} \l( 20.2 , 21.0 \r)$ & $\l( -1.1 , 1.9 \r) - \text{i} \l( 17.8 , 22.4 \r)$ \\
    $^{13}S_1$ & $\l( -7.1 , 28.8 \r) - \text{i} \l( 45.5 , 49.2 \r)$ & $\l( 5.6 , 7.7 \r) - \text{i} \l( 49.2 , 60.5 \r)$ \\
    $^{33}S_1$ & $\l( -17.6 , 7.0 \r) - \text{i} \l( 128.9 , 134.4 \r)$ & /~ \footref{note1} \\
\hline
\hline
\end{tabular}
}
\end{table}

\section{ Summary and Outlook}
\label{sec:summary}
We have studied the $\bar{N}N$ interaction in a hybrid approach, in which the contribution from the elastic process is determined in the LO covariant ChEFT, while the contribution from the annihilation process is described in the NLO HB ChEFT. The corresponding LECs were determined by fitting to the phase shifts and inelasticities provided by the PWA of the $\bar{p}p$ scattering data~\cite{Zhou:2012ui}. The overall description of the PWA with the hybrid  potential is comparable to that obtained with the NLO HB  potential. In addition, we searched for  near $\bar{N}N$ threshold structures, and found several bound states in the $^{11}S_0, ^{13}P_0, ^{13}S_1 $, and $^{33}S_1$ channels. The quantum number of $^{11}S_0$ supports the pseudoscalar interpretation of $X(1835)$, $X(1840)$, and $X(1880)$ observed by the BESIII Collaboration. However, the mass of this bound state is much smaller than $X(1835)$, $X(1840)$, and $X(1880)$. More detailed studies are needed to confirm the nature of this state. 

Although the $\bar{p}p$ data can be described reasonably well in the hybrid  approach, comparable to the NLO HB  results, further refinements can still be made. For example,  the potential responsible for the annihilation process is approximated in the conventional Weinberg power counting, the theoretical uncertainties are estimated roughly by varying the cutoff, and full renormalization group invariance has not been achieved. We will study these issues in the future.

\section{Acknowledgments}
This work was supported in part by the National Key R\&D Program of China under Grant No.2023YFA1606703, the National Natural Science Foundation of China under Grants No.12435007 and No.12347113, and the Chinese Postdoctoral Science Foundation under Grants No.2022M720360. Yang Xiao and Jun-Xu Lu thank the Fundamental Research Funds for Central Universities for the support. We thank Dr. Xian-Wei Kang for the enlightening discussions regarding the annihilation potential. Yang Xiao thanks Dr. Chun-Xuan Wang for the valuable discussions.

\appendix
\section{Generalized Fierz identities}\label{sec:Fierz}
This section briefly introduces the generalized Fierz identities; a detailed derivation can be found in Ref.~\cite{Nieves:2003in}. We start with some notations.
The Clifford algebra $\Gamma_i$ matrices are, 
\begin{align}
\nonumber
    \Gamma_S &= \mathbbm{1},\\\nonumber
    \Gamma_V &= \gamma_\mu,\\\nonumber
    \Gamma_T &= \sigma^{\mu\nu},\\\nonumber
    \Gamma_{AV} &= i\gamma^{\mu}\gamma_5,\\
    \Gamma_A &= \gamma_5.
\end{align}
An ordering of quadrilinears is defined as
\begin{align}
    e_I \l( 1234 \r) = \l(\bar{\Psi}_1 \Gamma_I \Psi_2 \r)\l(\bar{\Psi}_3 \Gamma^I \Psi_4 \r),
\end{align}
where $I \in  \{S,A,V,AV,T\}$. In this notation, the standard Fierz transformation gives the relation between the $e_I (1234)$ and the $e_J (1432)$,
\begin{align}
\label{eq:Fierz}
    e_I\l(1234\r) = \sum_J F_{IJ} e_J \l(1432 \r),
\end{align}
where $F_{IJ}$ is the matrix element of a $4 \times 4 $ matrix $\bm{F}$,
\begin{equation}
    \bm{F} = \frac{1}{4}\l(
    \begin{matrix}
        1 & 1 & \frac{1}{2} & -1 & 1 \\
        4 & -2 & 0 & -2 & -4\\
        12 & 0 & -2 &0 &12\\
        -4 & -2 &0 &-2& 4\\
        1& -1& \frac{1}{2} & 1 &1
    \end{matrix}
    \r).
\end{equation}
 Eq.~\eqref{eq:Fierz} can be abbreviated as
 \begin{align}
     \bm{e}\l(1234 \r) = \bm{F} \l(1432 \r).
 \end{align}
In the standard Fierz relation, the exchanged spinors remain the same type, i.e., a $u$-spinor/$v$-spinor remains a $u$-spinor/$v$-spinor. It is possible to interchange a pair of $u$-spinors to $v$-spinors in quadrilinears. For illustration, we consider a simple example where we
want to interchange the positions of the spinors in the first and second place. The results of other rearrangements can be obtained similarly.
Consider a quadrilinear
\begin{align}
    e_I \l( 2^c1^c34 \r) = \l( \bar{\Psi}^c \Gamma_I \Psi^c \r) \l( \bar{\Psi} \Gamma^I \Psi \r),
\end{align}
where $\Psi^c$ denotes that if $\Psi$ is a $u$-spinor, $\Psi^c$ is a $v$-spinor and vice versa. $\Psi^c$ and $\Psi$ are related by,
\begin{align}
    {\Psi}^c = \gamma_0 C  {\Psi}^*,
\end{align}
with $C$ the aforementioned charge transformation operator, and 
\begin{align}
\label{eq:charge}
    C^{-1}\Gamma_I C = \eta_I \Gamma_I^\T,
\end{align}
with the value of $\eta_I$ is
\begin{equation}
    \eta_I = \l\{ 
    \begin{matrix}
     +1 & I = S, AV, A\\
     -1 & I= V, T
    \end{matrix}
    \r.
    .
\end{equation}
Using Eq.~\eqref{eq:charge} and some matrix algebra, we obtain 
\begin{align}
     \bar{\Psi}  \Gamma_I \Psi = -\eta_I \bar{\Psi}^c \Gamma_I \Psi^c .
\end{align}
Therefore, we obtain the relation between the quadrilinears $e_I \l( 1234 \r)$ and the quadrilinears that the position of the first and second spinors are interchanged $e_J \l( 2^c1^c34 \r)$,
\begin{align}
    e_I \l( 1234 \r) =\sum_J S_{IJ} e_J \l( 2^c1^c34 \r),
\end{align}
where $S_{IJ}$ is the element of the matrix
\begin{align}
   \bm{S}= \text{diag} \l(-1, +1,+1,-1,-1 \r). 
\end{align}
Following the procedure introduced above and making full use of the standard Fierz transformations, we can obtain the generalized Fierz identities,
\begin{align}
\bm{e}\l(1234\r) = \bm{K}^{(abcd )} \bm{e}(abcd),
\end{align}
where the matrix $K$ is summarized in Table~\ref{tb:matrix_K}.
\begin{table}[h!]
\centering
\caption{Fierz matrices for all scalar combinations.}
\label{tb:matrix_K}
\begin{tabular}{l|l}
\hline
\hline
  Final order   & $\bm{K}$ \\
     \hline
   $(1234)$  & $\mathbbm{1}$ \\
   $(1432)$  & $\bm{F}$ \\
   $(2^c1^c34)$  & $\bm{S}$ \\
   $(124^c3^c)$  & $\bm{S}$ \\
   $(13^c2^c4)$  & $\bm{SFS}$ \\
   $(13^c4^c2)$  & $\bm{SF}$ \\
   $(142^c3^c)$  & $\bm{FS}$ \\
   $(2^c1^c4^c3^c)$  & $\bm{SS}=\mathbbm{1}$ \\
   $(31^c 2^c4)$  & $\bm{SF}$ \\
   $(31^c4^c 2)$  & $\bm{SFS}$ \\
   $(4^c 1^c 2^c 3 ^c)$  & $\bm{F}$ \\
   $(4^c1^c 32)$  & $\bm{FS}$ \\
\hline
\hline
\end{tabular}
\end{table}

\bibliography{Refs}

\end{document}